\documentclass[reprint,
superscriptaddress,
amsmath,amssymb,
aps,
twocolumn,
prb,
floatfix,
]{revtex4-1}

\usepackage{dcolumn}
\usepackage{amsmath,amssymb,graphicx,comment,bm,color,amsfonts,xcolor}
\usepackage{epsfig,epstopdf}
\usepackage{bm}
\usepackage{multirow}
\usepackage{wasysym}

\newcommand{\bs}{{\bf {s}}}

\newcommand{\bv}{{\bf {v}}}
\newcommand{\bk}{{\bf {k}}}

\newcommand{\etal}{{\it et al.}~}

\makeatletter
\newcommand{\thickhline}{%
    \noalign {\ifnum 0=`}\fi \hrule height 1pt
    \futurelet \reserved@a \@xhline
}
\newcolumntype{"}{@{\hskip\tabcolsep\vrule width 1pt\hskip\tabcolsep}}
\makeatother

\begin{document}

\title{Vorticity Locking and Pressure Dynamics in Finite-Temperature Superfluid Turbulence}
\author{Jason Laurie}
\email{j.laurie@aston.ac.uk}
\affiliation{Department of Mathematics, College of Engineering and Physical Sciences, Aston University, Aston Triangle, Birmingham, B4 7ET, United Kingdom}

\author{Andrew W. Baggaley} 
\email{andrew.baggaley@ncl.ac.uk}
\affiliation{Joint Quantum Centre (JQC) Durham--Newcastle, School of Mathematics and Statistics, Newcastle University, Newcastle upon Tyne, NE1 7RU, United Kingdom}


\date{\today}

\begin{abstract}
We present a numerical study of finite-temperature superfluid turbulence using the vortex filament model for superfluid helium. We examine the phenomenon of vorticity locking between the normal and superfluid components across a wide range of temperatures, using two different structures of external normal fluid drive.  Our analysis is restricted to one-way coupling between the two components, and subject to this simplification, we show that vorticity locking increases with temperature leading to the superfluid flow being more influenced by the characteristics of the normal fluid. This results in stronger superfluid polarization and  deviations from Gaussian statistics with a more probable occurrence of extreme fluctuations.  We also examine how these properties influence the pressure field and attempt to verify a long-standing $P_k\propto k^{-7/3}$ theoretical quantum signature within the spatial pressure spectrum.
\end{abstract}

\maketitle

\section{Introduction}
\noindent
Understanding the complex dynamics of finite-temperature superfluid turbulence is one of the upmost challenges for condensed-matter fluid dynamicists. There are a number of motivating factors for this, including recent experiments highlighting important connections between turbulence in quantum fluids and that of classical fluids~\cite{maurer_local_1998}, as well as regimes in which there are fundamentally different physics between the two~\cite{walmsley_dynamics_2014}. This spurs on a long-term hope that the discreteness of superfluid turbulence will help inspire breakthroughs of unsolved problems in classical turbulence theory. Moreover, we are discovering a growing number of physical systems where finite-temperature superfluid motion is believed to exist and is intrinsically important for understanding physical observations, these include neutron stars~\cite{andersson_superfluid_2007}, dark matter~\cite{berezhiani_theory_2015}, and also the concept of holographic duality providing a link between superfluid turbulence and the physics of black holes~\cite{chesler_holographic_2013,adams_holographic_2014}.

The two-fluid description~\cite{tisza_theory_1947,landau_theory_1941,landau_theory_1949} provides a theoretical framework for describing the remarkable two fluid system (consisting of a viscous normal fluid fully coupled to an inviscid superfluid component) observed in finite temperature superfluid helium and the turbulent motion it can undergo. There is a coupling between the two fluids through mutual friction, principally localized at the center of quantum vortices as well as through the overall pressure field. It is currently an experimental challenge to try and ascertain dynamics of each individual component through the limitations of experimental probes, such as second sound detection~\cite{skrbek_four_2000,babuin_quantum_2012}, hot-film or Pitot tubes~\cite{rousset_superfluid_2014,rusaouen_intermittency_2017}, pressure transducers \cite{rusaouen_detection_2017}, or tracking of tracer particles~\cite{paoletti_velocity_2008,gao_statistical_2017}. In this article we use state-of-the-art numerical simulations to probe finite-temperate superfluid turbulence in a way compatible with experimental measurements to discover new ways to interpret and analyze both fluid components of superfluid turbulence.

In a series of pioneering experiments in the 1950s, Vinen showed that turbulence in a superfluid can be generated in the laboratory by applying a heat flux to generate a counterflow between the normal and superfluid components~\cite{vinen_mutual_1957-2,vinen_mutual_1957-1,vinen_mutual_1957,vinen_mutual_1958}. Whilst Vinen's early experiments were the first to realize superfluid turbulence, the wider classical turbulence community's interest were piqued by the experimental results of Maurer and Tabeling~\cite{maurer_local_1998}, who investigated turbulence in helium-4 at temperatures ranging between $1.4~\textrm{K} -2.3~\textrm{K}$, i.e. both above and below the superfluid transition temperature $T_\lambda \approx 2.17~\textrm{K}$ under which superfluidity exists.  Even at temperatures where the superfluid fraction is above 90\% their results were consistent with Kolmogorov's 1941 theory prediction for the $k^{-5/3}$ energy spectrum of a 3D fluid. Subsequent experiments~\cite{salort_energy_2012} have confirmed that a similar agreement between normal and superfluid components is observed for higher-order statistics. What makes these results particularly interesting is that quantum mechanics constrains any rotational motion in the superfluid component to thin vortex lines or filaments, each carrying only a single unit of quantum of circulation $\kappa=9.97 \times 10^{-4} \mathrm{~cm}^{2} / \mathrm{s}$ in helium-4 meaning that vorticity is a discretized field in the superfluid component. Moreover, these vortex filaments are microscopic holes of the superfluid density of radius $a \approx 10^{-8} \mathrm{~cm}$ each emitting an irrotational inviscid fluid flow of velocity magnitude $v=\kappa /(2 \pi r)$ at a distance $r$ perpendicular from the axis of rotation.

The appearance of the same classical Kolmogorov energy spectrum in the superfluid component, known also as the quasi-classical Kolmogorov energy spectrum, should only occur within an inertial range of scales where the effects from large-scale external forcing or dissipation are negligible. This suggests we may observe a correlation between the velocity fields of the two components. Numerical results suggest this manifests itself in a locking of vorticity with the structure of the quantized vortex tangle directly mimicking structures in the normal fluid vorticity field~\cite{morris_vortex_2008}. The study of Morris~\etal~\cite{morris_vortex_2008} only considered a single temperature of $T=2.1~\textrm{K}$, where the superfluid component represents only $\approx 25\%$ of the total fluid density. More recent experimental~\cite{rusaouen_detection_2017} and theoretical~\cite{laurie_coarse-grained_2020} work has shown that local measurements of the pressure field can provide deep insights into the topological properties of the superfluid component and tangle therein. Indeed, the use of Pitot tubes~\cite{rusaouen_intermittency_2017} is a common strategy for experimentalists to extrapolate the superfluid velocity and high-order statistics in two-fluid experiments of superfluid helium. Motivated to revisit the work of Morris~\etal~\cite{morris_vortex_2008}, we investigate the phenomenon of vorticity locking across a wide range of superfluid temperatures paying careful attention to the characteristics of the pressure distribution in the system. We show that vorticity and velocity locking between the normal and superfluid components is present in simulations using both stationary and time-dependent normal fluid models. We further highlight the important role that temperature plays in this locking with correlations between the velocities of the two-components becoming weaker at low temperatures and particularly emphasized at small scales. When studying the pressure dynamics we confirm our previous theoretical work and the experimental study of Rusaouen~\etal~\cite{rusaouen_detection_2017}, which associated strong deviations from Gaussianity in the pressure field with coherent bundles of quantized vortices. Here we go further and suggest that with increasing temperature, deviations from Gaussianity become increasingly extreme due to greater organization of the quantized vortex tangle. Finally, we examine the pressure spectrum across our simulations and suggest that this could be used as a tool to investigate the structure of a tangle of quantized vortices. 
  
\section{Models of finite-temperature superfluid turbulence}

\noindent Finite-temperature superfluid turbulence can be modelled by a two-fluid description of Landau and Tisza~\cite{tisza_theory_1947,landau_theory_1941,landau_theory_1949}, where thermal excitations are described by a viscous normal fluid that is in coexistence with an inviscid superfluid, coupled through a mutual friction force. For scales larger than the mean inter-vortex distance between neighboring quantized vortices, a common phenomenological model has been the HBVK equations~\cite{hall_rotation_1956,bekarevich_phenomenological_1961}. The HBVK model utilizes the traditional fluid dynamics \textit{macroscopic} description of fluid flow (the Navier-Stokes equations) with each component  associated with separate velocity and density fields, denoted ${\bf v}_n$ and $\rho_n$ for the normal fluid and ${\bf v}_s$ and $\rho_s$ for the superfluid respectively, with total fluid density $\rho=\rho_{n}+\rho_{s}$ whose ratio is strongly temperature dependent, with $\rho_{n}/\rho_{s}\to 0$ in the zero-temperature limit. In the HVBK equations the normal fluid is modelled by the Navier-Stokes equations coupled via a mutual friction term $\mathbf{F}_{ns}$ to a coarse-grained inviscid superfluid component $\mathbf{v}_{s}$ modelled by the Euler equations:
\begin{subequations}
\begin{align}
\frac{\partial\mathbf{v}_{n}}{\partial t}+\left(\mathbf{v}_{n}\cdot\nabla\right)\mathbf{v}_{n}&=  -\frac{1}{\rho}\nabla P+\nu\nabla^{2}\mathbf{v}_{n}+\frac{\rho_{s}}{\rho}\mathbf{F}_{ns},  \label{eq:navier}\\
\frac{\partial\mathbf{v}_{s}}{\partial t}+\left(\mathbf{v}_{s}\cdot\nabla\right)\mathbf{v}_{s}&=  -\frac{1}{\rho}\nabla P-\frac{\rho_{n}}{\rho}\mathbf{F}_{ns},\label{eq:euler}\\
 \nabla\cdot {\bf v}_n=0, &\quad\nabla\cdot {\bf v}_s=0. 
\end{align}
\end{subequations}
Here, $P$ is the pressure field, $\nu$ is the kinematic viscosity of the normal fluid component, and ${\bf F}_{ns}$ is the mutual friction term that provides coupling between the normal and superfluid components and acts principally at the regions of high superfluid vorticity.

The HBVK model has shown success in probing the transition to turbulence in superfluid counterflow~\cite{melotte_transition_1998} and regimes of superfluid turbulence which do not support an energy cascade~\cite{barenghi_regimes_2016}. However, due to its coarse-grained description of the superfluid component it produces a continuous field representation of the superfluid vorticity that fails to resolve details close to or below the mean inter-vortex spacing $\ell$ (the average distance between neighboring quantized vortices).

An alternative \textit{mesoscopic} model for the superfluid velocity $\mathbf{v}_{s}$ was introduced by Schwarz~\cite{schwarz_three-dimensional_1985,schwarz_three-dimensional_1988} who considered the dynamics of one-dimensional vortex filaments through the vortex filament model (VFM). This can then be coupled to the Navier-Stokes equations (or indeed any analytic model) for the normal fluid component through an associated mutual friction term. 
The advantage of this method is that it  preserves the discreteness of the superfluid vorticity field and permits a description of the superfluid velocity at scales far below the inter-vortex spacing (unlike the HBVK equations) leading to a non-coarse-grained superfluid velocity field. 

In the VFM an evolution equation for the vortex filaments is given by the balance of the Magnus and drag forces acting on a vortex filament:
\begin{align}
\frac{{\rm d}{\bf s}}{{\rm d}t} & ={\bf v}_{s}+\alpha{\bf s}'\times({\bf v}_{n}-{\bf v}_{s})-\alpha'{\bf s}'\times\left[{\bf s}'\times\left({\bf v}_{n}-{\bf v}_{s}\right)\right],\label{eq:Schwarz-1}
\end{align}
where ${\bf s}(\xi,t)$ is the position of the one-dimensional space curves representing quantized vortex filaments in 3D Euclidean space. The mutual friction with the normal fluid is described by the last two terms governed by the non-dimensional temperature dependent friction coefficients $\alpha$ and $\alpha'$, ${\bf s}'={\rm d}{\bf s}/{\rm d}\xi$ is the unit tangent vector at the point $\mathbf{s}$, $\xi$ is arc length, and ${\bf v}_{n}$ is the normal fluid velocity at the point ${\bf s}$.

The velocity of the superfluid component ${\bf v}_{s}$ can be decomposed into a self-induced velocity generated by the vortex tangle ${\bf v}_{s}^{{\rm si}}$, and an external superfluid flow ${\bf v}_{s}^{{\rm ext}}$ such that ${\bf v}_{s}={\bf v}_{s}^{{\rm si}}+{\bf v}_{s}^{{\rm ext}}$. Here, the self-induced velocity ${\bf v}_{s}^{{\rm si}}$ of the vortex line at the point ${\bf s}$, is computed using the Biot-Savart law~\cite{saffman_vortex_1992}
\begin{align}\label{eq:BS-1}
{\bf v}_{s}^{{\rm si}}({\bf s},t) & =\frac{\kappa}{4\pi}\oint_{\cal L}\frac{({\bf r}-{\bf s})}{\left|{\bf r}-{\bf s}\right|^{3}}\times{\rm {\bf d}}{\bf r},
\end{align}
where the line integral extends over the entire vortex configuration $\mathcal{L}$. The external superfluid flow ${\bf v}_{s}^{\rm ext}$ is an externally imposed irrotational flow arising through either an excitation mechanism of the superfluid component or through the conservation of total mass of helium-4 in the presence of a mean normal fluid flow. In the VFM, the superfluid vorticity field is discretized through the 1D vortex filament description, and while a (resulting) continuous superfluid velocity field can be recovered via Eq.~\eqref{eq:BS-1}, it can contain spurious values arising from the discretization of the filaments. 

The co-action of the two velocity components can most predominantly be observed through the fluid pressure $P$. The relationship between the pressure and the two fluid components can be determined through the two vorticity fields. This can more readily be shown by taking the divergence of the HVBK equations and using incompressibility of the fluid flow. One can relate the pressure $P$ to the vorticity through a Poisson equation involving  the spin tensor $\mathbf{W}_{i}= \left[\nabla\mathbf{v}_{i}-\nabla\mathbf{v}_{i}^{T}\right]/2$  and the strain tensor $\mathbf{E}_{i}= \left[\nabla\mathbf{v}_{i}+\nabla\mathbf{v}_{i}^{T}\right]/2$ for each superfluid and normal components, $i=s,n$ respectively
\begin{align}\label{eq:pressure}
\nabla^{2}P&=  \frac{\rho_{s}}{2}\left(\mathbf{W}_{s}:\mathbf{W}_{s}-\mathbf{E}_{s}:\mathbf{E}_{s}\right)\nonumber\\
&+\frac{\rho_{n}}{2}\left(\mathbf{W}_{n}:\mathbf{W}_{n}-\mathbf{E}_{n}:\mathbf{E}_{n}\right).
\end{align}
Here, $:$ denotes the double dot product between two matrices defined as $A:B = a_{ij}b_{ij}$ with index summation implied for the matrix elements $a_{ij}$ and $b_{ij}$. The contributions arising from the spin tensors is related to the vortical motion of the fluid components proportional to their respective fluid densities.  Terms involving the strain tensor lead to oppositely signed contributions determined by the local stretch and strain of the fluid flow. What is most evident is that in high vorticity regions the right-hand side of Eq.~\eqref{eq:pressure} will be overwhelmingly positive. Consequently, the pressure field (both temporally and spatially) contains important information on the distribution of vorticity across scales. 

\section{Our Numerical Approach}

We perform numerical simulations using the vortex filament method to study the correlation between the normal fluid and superfluid vorticity fields, and subsequently the pressure field dynamics across a range of experimentally relevant temperatures. We use the model of Schwarz, Eq.~\eqref{eq:Schwarz-1}, with no external superfluid flow ${\bf v}_{ext}$. Our calculations are performed in a periodic cube of width $D=0.1~\rm cm$. The numerical technique to which vortex lines are discretized into a number of points ${\bf s}_j$ for $j=1, \cdots N$ held at a minimum separation $\Delta\xi/2$, compute the time evolution, de-singularize the Biot-Savart integrals, evaluate ${\bf v}_s$, and algorithmically perform vortex reconnections when vortex lines come sufficiently close to each other, are described in detail in previous papers~\cite{baggaley_tree_2012,baggaley_sensitivity_2012}. The Biot-Savart integral is computed using a tree-algorithm approximation~\cite{baggaley_tree_2012} with opening angle set to $\theta=0.2$. We take $\Delta \xi=2.5\times 10^{-3}~\rm cm$ and apply a time step of $5\times 10^{-5}~\rm s$ for our time-integration. 

Ideally, we would consider a two-way coupling between the normal and superfluid components. Whilst a few studies have started to look at fully coupled two-fluid turbulence~\cite{kivotides_spreading_2011, yui_three-dimensional_2018,galantucci_new_2020}, these simulations are still extremely computationally expensive, limiting the size of the vortex structures and parameter space that can be studied. Moreover, there is a precedent for studying finite temperature superfluid turbulence with just a coupling from the normal to superfluid component~\cite{adachi_steady-state_2010,baggaley_vortex-density_2012,kondaurova_structure_2014} that will permit a direct comparison to previous studies, indeed this is precisely the strategy used in~\cite{morris_vortex_2008}. It is important, however, to add the caveat that whilst this may be more appropriate at the high temperatures considered in Ref.~\cite{morris_vortex_2008}, as we move to lower temperatures (and thus increasing superfluid fraction) the validity of this model assumption becomes more questionable.

In many experimental studies of finite-temperature superfluid turbulence, the large-scale flow is generated mechanically, for example by pushing helium through pipes or channels~\cite{salort_energy_2012}, using plungers, bellows or by stirring it with grids~\cite{smith_decay_1993}, or via propellers~\cite{maurer_local_1998}. Away from any boundaries it is reasonable to expect the normal fluid to be in a state of homogeneous, isotropic turbulence. We follow our previous work~\cite{sherwin-robson_local_2015} and consider two different models for this turbulent normal fluid velocity field. Firstly, we take a well established analytic model of a turbulent-like flow, comprised of the summation of random Fourier modes with an imposed Kolmogorov energy spectrum~\cite{osborne_one-particle_2005}, commonly referred to as the Kinematic Simulation (KS) model. More explicitly, we take

\begin{align}\label{eq:KS}
{\bf v}_n({\bf s},t)&=\sum_{m=1}^{M}\left[({\bf A}_m \times {\bf k}_m) \cos \left(\phi_m\right)
+({\bf B}_m \times {\bf k}_m) \sin\left(\phi_m\right)\right],
\end{align}

\noindent
where $\phi_m=\bk_m \cdot \bs + f_m t$, $\bk_m$ are wavevectors and $f_m=\sqrt{k^3_m E(k_m)}$ are angular frequencies.  The random parameters ${\bf A}_m$,  ${\bf B}_m$  and ${\bf k}_m$ are chosen so that the normal fluid's energy spectrum obeys Kolmogorov's scaling $E(k_m)\propto k_m^{-5/3}$ in the inertial range $k_1<k< k_M$, where $k_1 \approx 2 \pi/D$ and $k_M$ correspond to the outer scale of the turbulence and the dissipation  length scale respectively; we can readily define the Reynolds number through ${\rm Re}=(k_M/k_1)^{4/3}$. The synthetic turbulent flow defined by Eq.~\eqref{eq:KS} is solenoidal, time-dependent. Whilst it is not a solution of the Navier-Stokes equation numerical studies of $N$-particle dispersion compare well with statistics obtained in experiments and direct numerical simulations of the Navier-Stokes equation. With this being said, it does lack the coherent vortical structures, which are an intrinsic feature of Navier-Stokes turbulence~\cite{she_intermittent_1990}; this motivates consideration and comparison to an alternative model for the normal fluid's velocity field.

Thus, results using this approach will be directly compared to a normal fluid profile $\mathbf{v}_n$ obtained from a numerical snapshot of classical homogeneous and isotropic turbulence generated by the Navier–Stokes equations taken from the John Hopkins Turbulence Database~\cite{li_public_2008}. The data set consists of a velocity field discretized on a mesh of $1024\times 1024\times 1024$ points. The estimated Reynolds number of the velocity snapshot is $\operatorname{Re} \sim\left(L_{0} / \eta_{0}\right)^{4 / 3} \simeq 3025$, where $L_{0}$ and $\eta_{0}$ are the integral and Kolmogorov scales, respectively. The reason for using a single stationary snapshot for the normal fluid velocity profile and not a time-dependent one is based solely on computational constraints. As will be reported later in this paper, the stationarity of the DNS normal fluid flow does have an observable impact on the mixing between the two fluid components, however, we do not believe that it produces any nonphysical effects.

As the vortex filament method described above is a Lagrangian description, we are required to define the normal fluid velocity at the location of the discretization points $\bs_j$. For the KS model this is straightforward from the analytic form of $\bv_n$, for the DNS field we must interpolate the velocity at each discretization points, and we do this using a trilinear scheme. Moreover, in order to study the pressure field of the superfluid we must move in the other direction and move from a Lagrangian to Eulerian frame of reference. To do this we apply the following prescription to coarse-grain the velocity field onto an Eulerian mesh: First, we generate the superfluid velocity field $\mathbf{v}_{s}$, and sample both the superfluid ${bf v}_s$ and normal fluid $\bv_n$ velocity fields on a uniform three-dimensional spatial mesh of size $128\times 128\times 128$, using the tree-approximation to the Biot-Savart integral, Eq.~\eqref{eq:BS-1} for $\bv_s$ and sub-sampling in the case for $\bv_n$.  We then coarse-grain these fields by applying a Gaussian low-pass filter $G$ to each velocity component
\begin{align}\label{eq:filter}
\bar{\mathbf{v}}_i({\bf x},t) = \int  \mathbf{v}_i({\bf x}',t)G({\bf x}-{\bf x}')\ d{\bf x}',
\end{align}
 where $i=n,s$ and the Gaussian low-pass filter is defined through a filter scale $l_f$ as $G({\bf x}) = (2\pi l_f^2)^{-3/2}\exp(-|{\bf x}|^2/2l_f^2)$. The definition of the filter convolution in Eq.~\eqref{eq:filter} implies that one can simply multiply the Fourier harmonics of each velocity field by $\hat{G}_{\bf k} = \exp(-|{\bf k}|^2l_f^2/2)$ where we define the Fourier transform through $G({\bf x}) = (1/2\pi)^{3}\sum_{\bf k} \hat{G}_{\bf k}\exp(i{\bf k}\cdot {\bf x})$. In our previous work Ref.~\cite{laurie_coarse-grained_2020} we demonstrated that taking a filtering scale of $l_f=2\ell$ was sufficient to remove any discretization effects associated to the VFM whilst keeping enough detail to resolve important turbulent statistics close to the mean inter-vortex scale $\ell$. 


\section{Results}
\noindent We conduct a series of numerical simulations leading to statistically stationary finite-temperature superfluid turbulence across four different temperatures $T=1.3~{\rm K}, 1.5~{\rm K},2.0~{\rm K},2.15~{\rm K}$ (from $\rho_s/\rho \approx 96\% -  13\%$ respectively), using two models of the normal fluid flow: KS and the DNS snapshot described above.  
We initialize the system with a few seed vortex rings and allow the simulations to reach statistical steady states as measured by the total vortex line length $|\mathcal{L}|$. In each simulation, the strength of the normal fluid flow (drive) is adjusted to ensure that the final steady state configurations have comparable superfluid vortex line density of $L = |\mathcal{L}|/D^3 \approx 2\times 10^4 ~{\rm cm}^{-2}$, see Table~\ref{tab:1} for specific details regarding simulation diagnostics. We purposely chose each simulation to have equivalent mean inter-vortex spacing $\ell=1/\sqrt{L}~{\rm cm}$ in order to allow for a like-for-like comparison across temperature and types of normal fluid drive. Whilst this process ensures that the superfluid Reynolds number ${\rm Re_{s}} = v^s_{\rm RMS}L_0/\kappa$~\cite{volovik_classical_2003} remains comparable between simulations it does indeed change the Reynolds number of the normal fluid component. All of our subsequent analysis is performed within statistically steady state regimes, including the computation of the superfluid and normal fluid velocities and application of the coarse-grain procedure described above on a $128\times 128\times 128$ uniform spatial grid.  We have compared the presented results with those at alternatives time in steady state conditions and observed no observable difference in statistics. Therefore, we are confident that we are not statistically sampling our simulations during any extreme turbulent event.

\begin{table*}[htp!]
\caption{Numerical measurements of the final turbulent states in the eight simulations we perform. Four forced by a static Navier-Stokes rendered normal fluid flow (DNS) and four using a kinematic simulation (KS) normal flow. Each set is simulation at temperatures $T=1.3, 1.5, 2.0, 2.15~\rm K$. \label{tab:1}}
{\renewcommand{\arraystretch}{1.4}%
\begin{tabular}{ |c"c|c|c|c"c|c|c|c| }
\hline
& \multicolumn{4}{ c" }{DNS} & \multicolumn{4}{ c| }{KS}\\
\hline
 & $1.3~\rm K$ & $1.5~\rm K$ &$2.0~\rm K$&$2.15~\rm K$ & $1.3~\rm K$ & $1.5~\rm K$ &$2.0~\rm K$&$2.15~\rm K$  \\ \hline
 $\alpha  $ & $0.0340$ & $0.0720$ &$0.279$&$0.893$ & $0.0340$ & $0.0720$ &$0.279$&$0.893$  \\ \hline
 $\alpha' $ & $0.0138$ & $0.0177$ &$0.0120$&$-0.192$ & $0.0138$ & $0.0177$ &$0.0120$&$-0.192$  \\ \thickhline
$|\mathcal{L}| ~\rm cm$& $20.746$ & $20.581$ & $21.467$& $20.382$ & $21.490$ & $20.881$ & $21.867$ & $22.099$ \\ \hline
$\ell~\rm cm$& $0.00694$ &  $0.00697$ &    $0.00683$ &   $0.00700$ & $0.00682$ &  $0.00676$ & $0.00676$ & $0.00673$\\ \hline
$v^n_{\rm RMS}~\rm cm\,s^{-1}$  & $1.225$ & $0.787$ & $0.426$ & $0.341$ & $2.057$ & $1.322$ & $0.624$ & $0.404$ \\ \hline
$v^s_{\rm RMS}~\rm cm\,s^{-1}$ & $0.336$ & $0.264$ & $ 0.246$ & $0.298$ & $0.383$ & $0.324$ & $0.282$ & $0.291$ \\ \hline
$\omega^n_{\rm RMS}~\rm s^{-1}$ & $185.972$ & $130.050$ & $70.444$  & $56.355$ & $363.860$ & $233.909$ & $110.458$ & $71.473$ \\ \hline
$\omega^s_{\rm RMS}~\rm s^{-1}$ & $31.139$ & $25.859$ & $26.080$  & $31.293$ & $38.669$& $27.906$ & $25.370$ & $24.955$ \\
\hline
$P_{\rm RMS}~\rm kg\,cm^{-1}\,s^{-2}$ & $0.218$ &  $0.238$ & $0.321$ & $0.314$& $0.460$ & $0.474$& $0.513$& $0.338$ \\ \hline
\end{tabular}
}
\end{table*}

\subsection{Polarized and Random Vortex Line Density}

\begin{figure}[htp!]
\includegraphics[width=0.9\columnwidth]{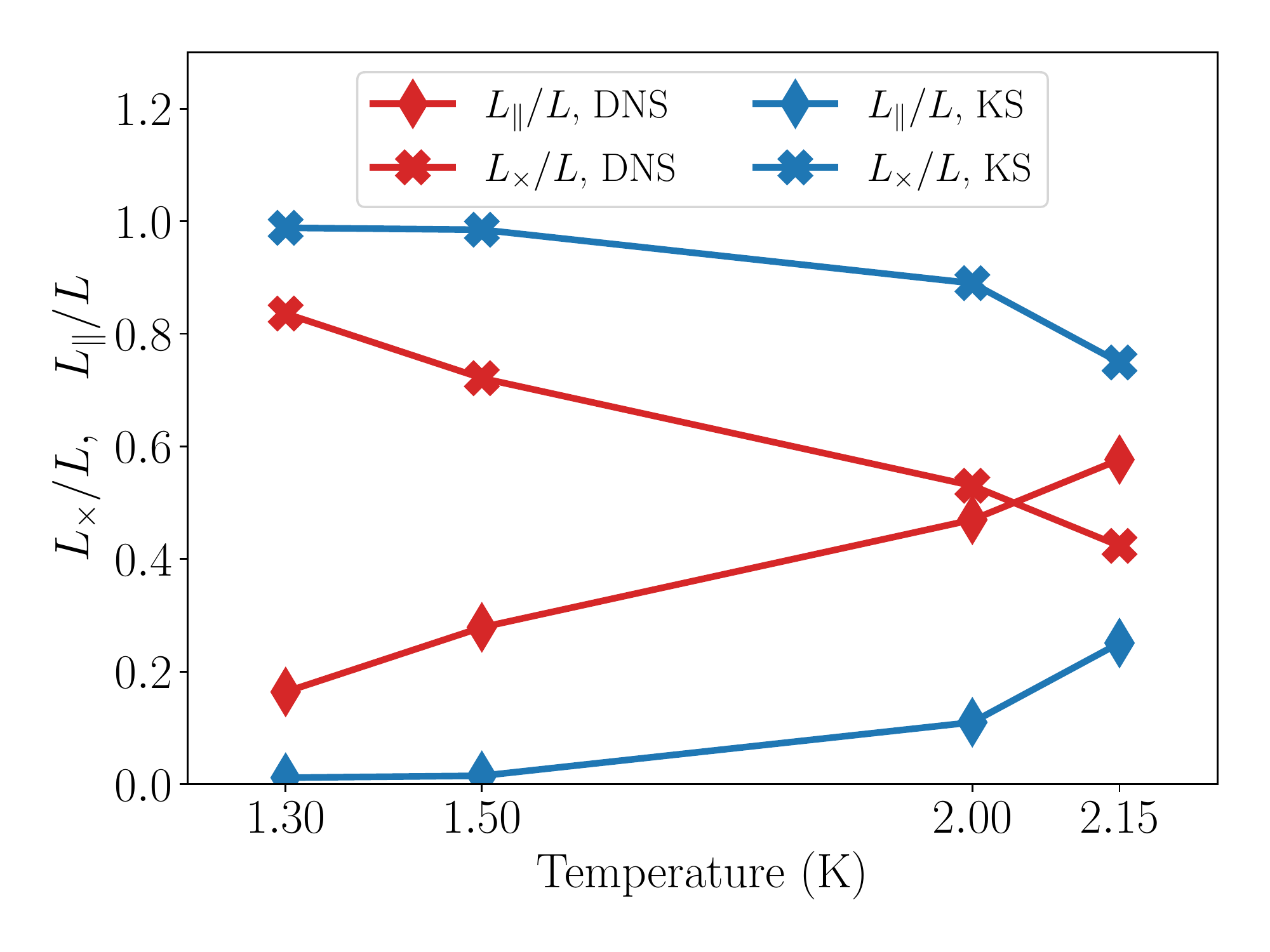}
\caption{Proportions of the polarized $L_{\parallel}$ and random $L_\times$ vortex line densities of our steady state vortex configurations across our simulations. In all cases, the polarized vortex line length is defined as those vortex line elements with a local polarization  larger than the RMS values of the $1.5~\rm K$ DNS simulation. \label{fig:polarization}}
\end{figure}

\noindent A first hint at the role of temperature in the correlations between the two fluid components motion can be observed in Table~\ref{tab:1}. We observe negative temperature dependence of the normal fluid root-mean-square (RMS) velocities due to the need to adjust the normal fluid drive with respect to temperature in order to have comparable superfluid vortex line densities. What is particularly interesting is that we do observe some fluctuations across the superfluid velocity and vorticity RMS values with respect to temperature that appear uncorrelated to temperature for the DNS simulations but negatively correlated in the case of the normal KS flow. Due to the similar total superfluid vortex line densities, which is typically used as a proxy for superfluid kinetic energy, we would expect little variation. However, we are likely observing natural temporal fluctuations in the flow statistics due to our data being produced from spatial averages only at a fixed moment in time. With that being said, the structure of the superfluid flow could have a significant impact on the observed statistics. Barenghi and Roche~\cite{roche_vortex_2008} argued that $L$ can be decomposed into a polarized field (carrying the majority of the fluid kinetic energy), i.e. in the form of coherent structures such as vortex bundles~\cite{baggaley_vortex-density_2012}, and an isotropic randomly orientated field of vortex lines passively advected. Such a picture is common in the literature of magneto-hydrodynamics, and in particular mean-field dynamo theory~\cite{krause_mean-field_2016} where more formally one can decompose the vortex line density $L$ as
\begin{align} \label{eq:Ldecomp}
L=L_{\parallel}+L_\times,
\end{align}
where $L_{\parallel}$ denotes the polarized component and $L_\times$ the randomly orientated component. In Fig.~\ref{fig:polarization} we plot the proportions of the polarized and random components of the vortex line density across our simulations. We use the same algorithm as detailed in Ref.~\cite{baggaley_vortex-density_2012} to compute a \textit{local vorticity} for each vortex line segment on the superfluid vortex tangle by locally averaging the superfluid vorticity field $\boldsymbol{\omega}_{s}$ at the discretization points using an $M_4$ cubic spline kernel of finite support~\cite{monaghan_smoothed_1992}. This approach allows us to compute a local vorticity vector $\boldsymbol{\omega}^{\rm loc}$ on the discretized superfluid vortex tangle (as opposed to the uniform mesh in which $\bar{\boldsymbol{\omega}}_s$ is defined). Specifically, we define the local vorticity at each segment by
\begin{align*}
\boldsymbol{\omega}^{\rm loc}({\bf s}_i)=\kappa \sum_{j=1}^N {\bf s}_j^\prime W(r_{ij},h)\Delta \xi_j,
\end{align*}
where $r_{ij}=|{\bf s}_i - {\bf s_j}|$, $W(r,h)=g(r/h)/(\pi h^3)$, and
\begin{align*}
g(q)=\begin{cases}  1- \frac{3}{2}q^2 + \frac{3}{4}q^3 & \text{for $0 \leq q < 1$},\\
  \frac{1}{4}(2-q)^3 & \text{for $1\leq q < 2$},\\
  0 & \text{otherwise}.
\end{cases}
\end{align*}
The radius of support of the kernel is chosen to be the mean inter-vortex spacing $h=\ell$ to ensure sufficient local averaging to capture polarization of the vorticity distribution from coherent structures.

Using the tangle averaged value of $\boldsymbol{\omega}^{\rm loc}$ from the $T=1.5~\rm K$ DNS simulation as the threshold value for comparison across all simulations, the polarized vortex line segments are then classified as those with values larger than this threshold value and are subsequently assigned to $L_\parallel$. Segments with local vorticity below the threshold are attributed to $L_\times$. This enables us to compare the relative proportion of polarization defined by a fixed threshold value. We observe a clear temperature dependence of the vortex line density polarization, with increasing local polarization with temperature. Furthermore,  the DNS simulations indicate enhanced polarization when compared across the simulations of the KS normal flow. Indeed, this is to be expected; for higher temperatures, the mutual friction between the two fluid components is stronger meaning that there is an increased coupling between the two components, more vorticity locking (see subsection below), with small scale fluctuations on the quantum vortex lines (Kelvin waves) more readily dissipated. Consequently, at higher temperatures, the superfluid vortex tangle would be smoother and increasingly correlated to the background large-scale normal fluid flow. In this respect, as studied in classical turbulence~\cite{she_intermittent_1990}, the uncorrelated and random phases of the KS flow will lead to reduced spatial coherence of the resulting physical space velocity field and subsequently less extreme vortical structures when compared to Navier-Stokes generated DNS (see Fig.~\ref{fig:tangle} for a visual verification).

\subsection{Vorticity Locking}
\begin{figure*}[htp!]
\begin{center}
\includegraphics[width=0.9\textwidth]{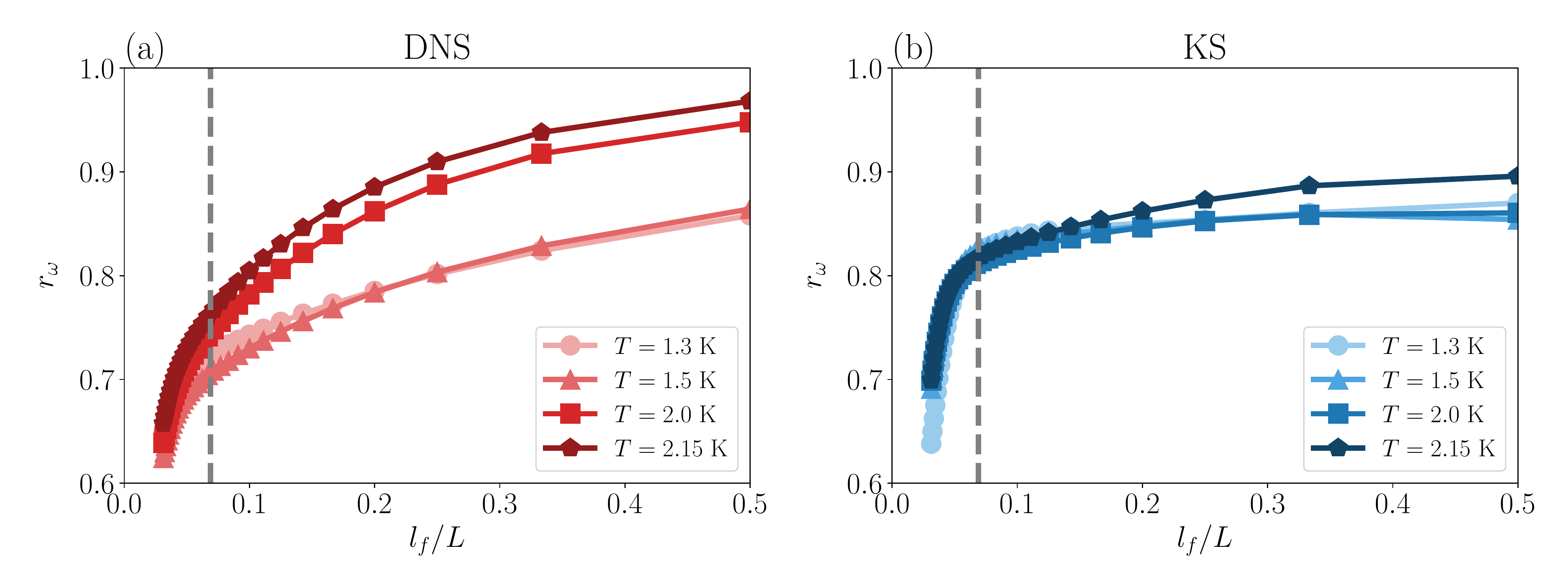}
\caption{Vorticity magnitude correlation $r_\omega$ versus filtering scale $l_f/L$ for the DNS (left) and KS (right) simulations. The gray dashed vertical lines indicate the scale of the mean inter-vortex spacing $\ell$ which is similar over all temperatures and normal fluid types. \label{fig:koplik}}
\end{center}
\end{figure*}

As just mentioned, for finite-temperature superfluids, the normal and superfluid components are coupled through mutual friction, described by the force ${\bf F}_{ns}$ in our model. The exact form of this expression in the momentum equation is proportional to the difference between the two fluid velocities, namely $\propto \left({\bf v}_n - {\bf v}_s\right)$, specified by temperature dependent coefficients $\alpha$ and $\alpha'$, acting close to regions of high superfluid vorticity. Hence, for non-zero temperatures,  the mutual friction force will act to bring the local velocity and vorticity fields back towards that of the other component predominately in regions close to the center of the vortices, in essence locally locking  both components of the velocity and vorticity fields together. If both velocity fields are fully locked then the mutual friction force vanishes. Following the approach of Morris~\etal~\cite{morris_vortex_2008} we can quantitatively measure vorticity locking by defining a vorticity correlation coefficient of the form
\begin{align}\label{eq:r}
r_\omega\left(l_f\right)&=\frac{\left\langle \bar{\omega}_{s} \bar{\omega}_{n}\right\rangle}{\sqrt{\left\langle \bar{\omega}_{s} ^{2}\right\rangle\left\langle \bar{\omega}_{n}^{2}\right\rangle}},
\end{align}
using spatial averages between the coarse-grained superfluid and normal fluid vorticity fields. Here $\bar{\omega}_{i} = |\bar{\boldsymbol{\omega}}_{i}|$ denotes the coarse-grained vorticity magnitude, coarse-grained at a scale $l_f$, for $i=n,s$ and $\langle \cdot \rangle$ denoting a spatial average over the 3D domain. Our measurements of the vorticity correlation across temperature and normal fluid form are presented in Fig.~\ref{fig:koplik}, and can be directly compared to those of Ref.~\cite{morris_vortex_2008} who found $r_\omega \approx 0.6$ at $T=2.1~\rm K$ at a filtering scale $l_f=\ell$. Our correlation coefficients for both normal fluid velocity fields are marginally larger, which could be a consequence of a different coarse-graining procedure, or in the case of the DNS simulation, the fact that the turbulence is frozen in time which would plausibly inflate the value of $r_\omega$ due to the lack of temporal motion. With that being said, we can draw the same insight as Morris~\etal~\cite{morris_vortex_2008} and observe that at scales larger than the inter-vortex spacing, $\ell$, the normal and superfluid vorticity fields  become increasingly correlated. 

We can go further and note that the correlation is highly dependent on temperature, due to the enhanced mutual friction via the temperature-dependent coefficients. In particular, at lower temperatures, around the filtering scale of the mean inter-vortex spacing $l_f \approx \ell$, the correlation coefficient undergoes a sharp transition and drops off rapidly towards small filtering scales due to the likely influence of Kelvin waves along the quantized vortex lines randomizing the superfluid vorticity at such small scales. This provides further supportive evidence of our recent study on coarse-grained statistics in superfluid turbulence~\cite{laurie_coarse-grained_2020}, which demonstrated that a filtering scale close to $l_f=2\ell$ is the most appropriate length scale to produce a continuous superfluid  vorticity distribution from the discrete vortex filament model as this filters out the discreteness of the vortex filaments while preserving the large-scale coherent structures (of more than two vortex lines and above) and related velocity statistics.

\begin{figure}[htp!]
\begin{center}
\includegraphics[width=\columnwidth]{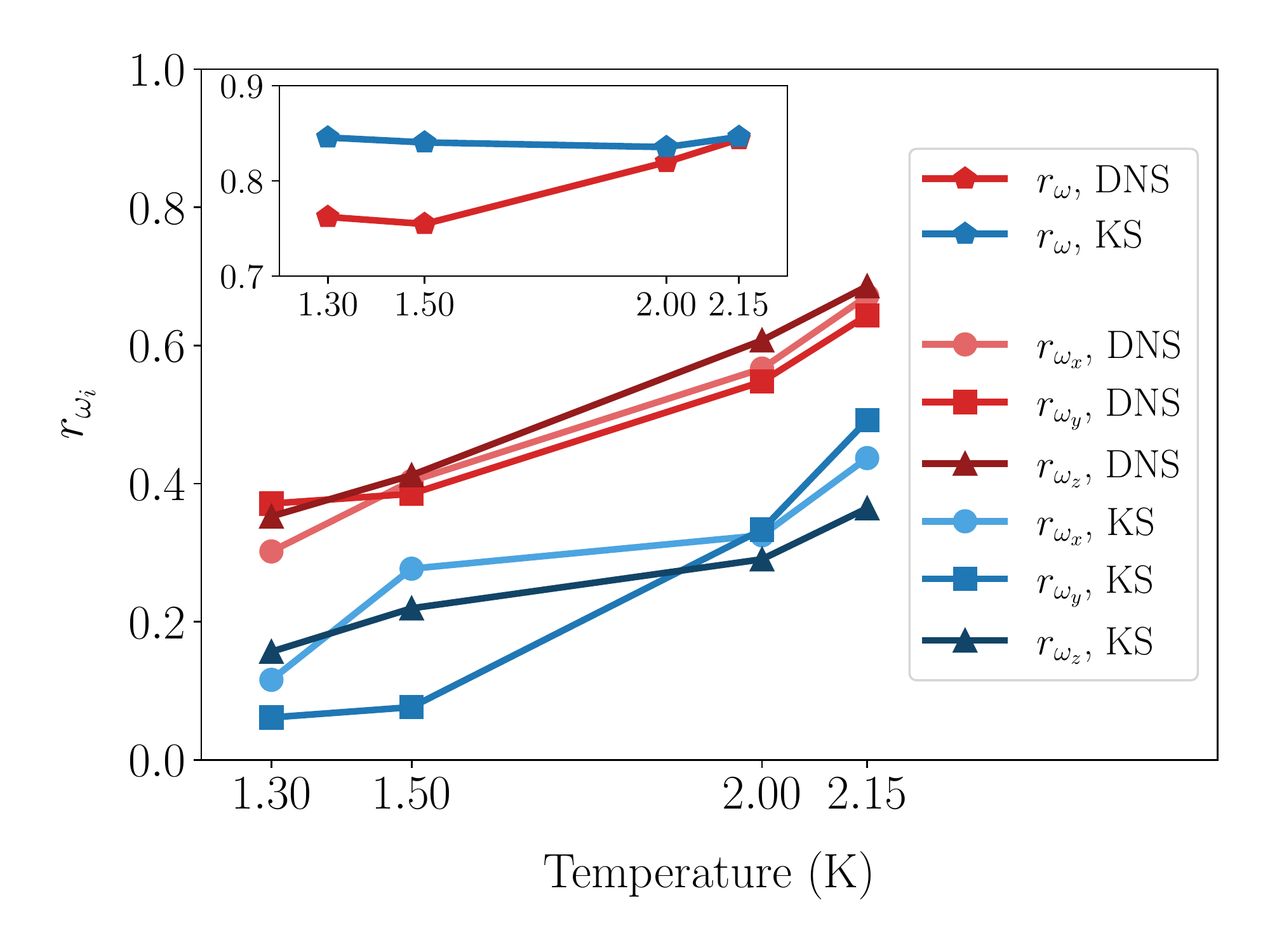}
\caption{Vorticity correlations versus temperature measured with a coarse-grained scale of $l_f=2\ell$. Inset shows the vorticity magnitude correlation $r_\omega$, Eq.~\eqref{eq:r}, while the main figure displays vorticity component correlation $r_{\omega_i}$, Eq.~\eqref{eq:r_xyz}.\label{fig:correlation}}
\end{center}
\end{figure}

We also examine the correlation of the individual components of the vorticity field, defined through the correlation coefficients
\begin{align}\label{eq:r_xyz}
r_{\omega_i}&=\frac{\left\langle \left(\hat{\mathbf{e}}_i \cdot \bar{\boldsymbol{\omega}}_{s}\right) \left(\hat{\mathbf{e}_i} \cdot\bar{\boldsymbol{\omega}}_{n}\right)\right\rangle}{\sqrt{\left\langle \left(\hat{\mathbf{e}}_i \cdot \bar{\boldsymbol{\omega}}_{s}\right)^2\right\rangle\left\langle \left(\hat{\mathbf{e}}_i \cdot \bar{\boldsymbol{\omega}}_{n}\right)^2\right\rangle}},
\end{align}
for $i=z,y,z$. In Fig.~\ref{fig:correlation} we plot the directional vorticity component correlation coefficient $r_{\omega_i}$, $i=x,y,z$, at scale $l_f=2\ell$ which reveals a more complex picture. First, we observe no obvious preference for direction which should be the case as all our simulations are for homogeneous and isotropic turbulence. Second, we see an almost linear temperature dependence of the component correlations for both sets of simulation data. In fact, the DNS simulations lead to uniformly more correlation across the temperature range probably due to the stationarity of the normal fluid flow. Finally, we observe that the correlation between the two fluid components is less for each individual spatial directions as opposed to the vorticity magnitudes (see insert of Fig.~\ref{fig:correlation}). This highlights that there is more variability between the vorticity directions than in the actual vorticity magnitudes. This is a reasonable observation when you consider that the turbulence is driven by either a fixed, in magnitude and phase, normal flow (DNS) or a dynamic normal flow (KS) prescribed only by deterministically varying phases in time, see Eq.~\eqref{eq:KS}. 

\begin{figure*}[htp!]
\begin{center}
\includegraphics[width=0.9\textwidth]{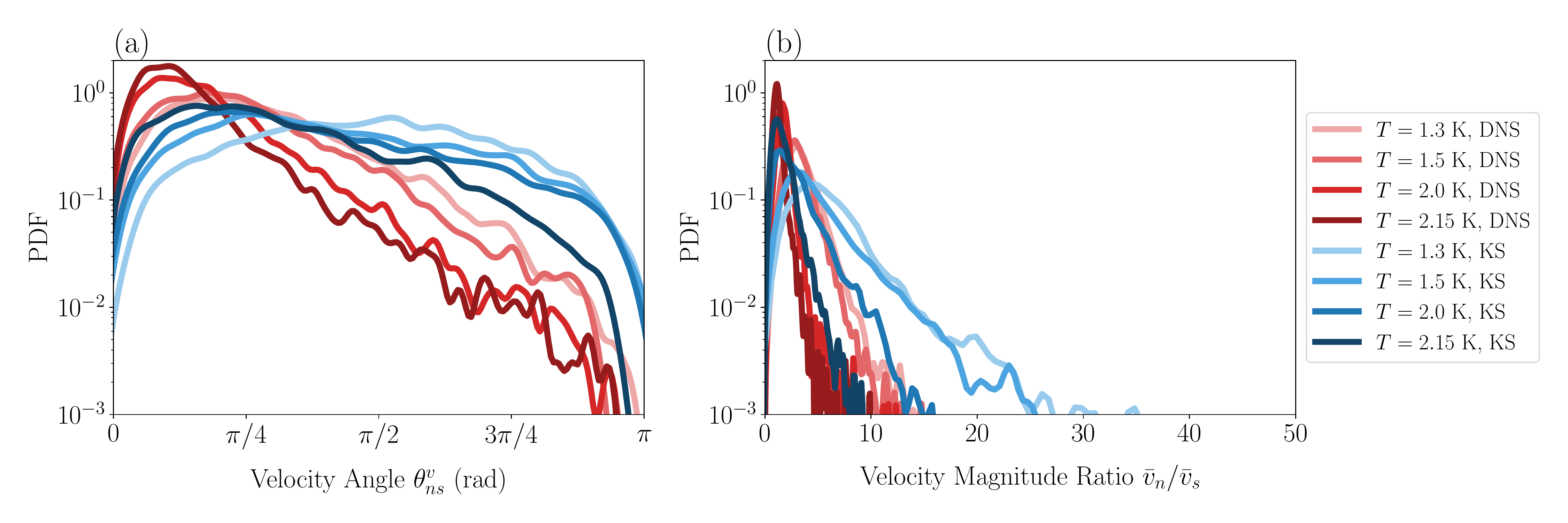}
\caption{ Probability density functions (PDFs) of the angle (left) and magnitude ratio (right) between the coarse-grained velocity fields $\bar{\mathbf{v}}_n$ and $\bar{\mathbf{v}}_s$  with respect to temperature and normal fluid drive.  \label{fig:vel_angles}}
\end{center}
\end{figure*}

\begin{figure*}[htp!]
\begin{center}
\includegraphics[width=0.9\textwidth]{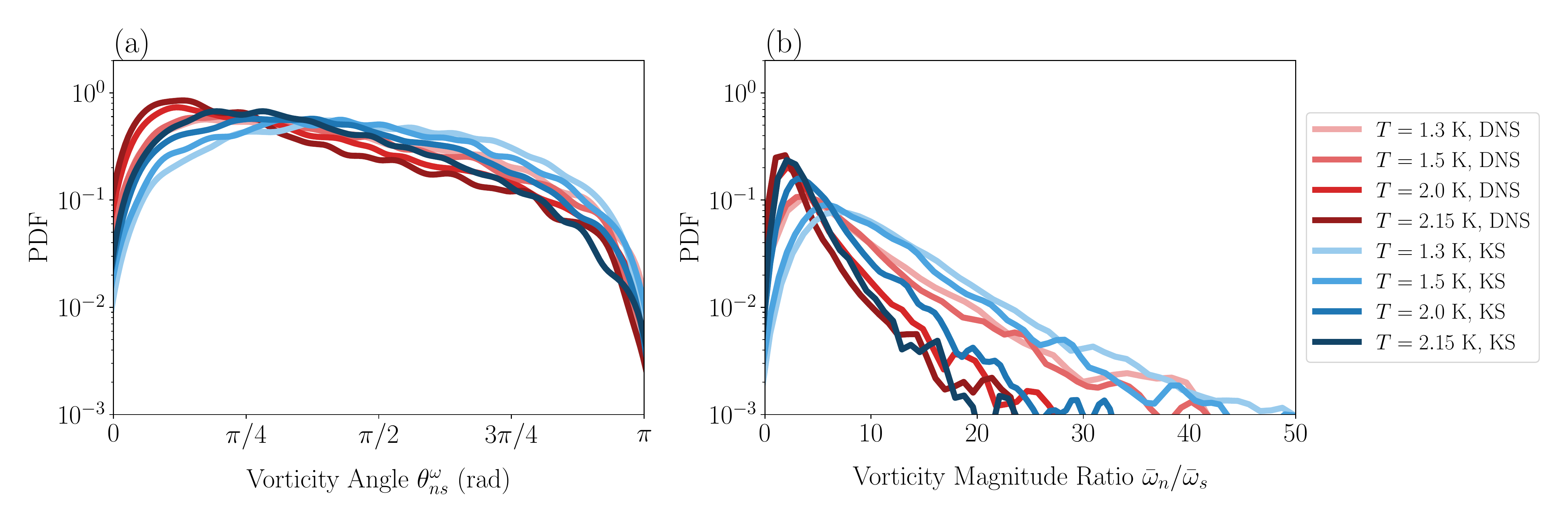}
\caption{
Probability density functions (PDFs) of the angle (left) and magnitude ratio (right) between the coarse-grained vorticity fields $\bar{\boldsymbol{\omega}}_n$ and $\bar{\boldsymbol{\omega}}_s$ with respect to temperature and normal fluid drive. 
 \label{fig:vort_angles}}
\end{center}
\end{figure*}

Whilst our results build on those of Ref.~\cite{morris_vortex_2008} and show a high degree of correlation between the components, particularly at larger scales, to demonstrate true locking between the velocity and vorticity fields we must go further. To do this we analyze the angle between the normal and superfluid velocity and vorticity fields, defined through
\begin{equation}
\theta_{ns}^v=\arccos \left( \frac{\bar{\mathbf{v}}_n \cdot  \bar{\mathbf{v}}_s}{\bar{v}_n \bar{v}_s} \right),\: \theta_{ns}^\omega=\arccos \left( \frac{\bar{\boldsymbol{\omega}}_n \cdot  \bar{\boldsymbol{\omega}}_s}{\bar{\omega}_n \bar{\omega}_s} \right),
\end{equation}
and the ratio of the magnitude of the two components, i.e. $\bar{v}_n/\bar{v}_s$,   $\bar{\omega}_n/\bar{\omega}_s$. 
Figures~\ref{fig:vel_angles} and \ref{fig:vort_angles} display these statistics over the suite of simulations we performed. Both show a similar trend with the PDFs of $\theta_{ns}^v$ and $\theta_{ns}^\omega$ becoming increasingly distributed around small angles (associated to positively aligned vectors) with increasing temperature, while the vector magnitude ratio PDFs of $\bar{v}_n/\bar{v}_s$ and $\bar{\omega}_n/\bar{\omega}_s$ become strongly peaked around unity. This is a clear indication of increased locking between the two components as temperature increases. That this is more pronounced in the velocity field is potentially a result of the mutual friction directly acting to minimize the difference between the two velocity fields.  Furthermore, even with the coarse-graining, an artifact of the discrete nature of the superfluid vorticity field leads to it becoming vanishingly small away from the quantized vortices that could result in reduced correlation in the vorticity field. These results, along with the correlation coefficients presented earlier, point to a pronounced locking of the velocity and vorticity fields as the temperature of the system increases. 


\begin{figure*}[t!]
\begin{center}
\includegraphics[width=0.8\textwidth]{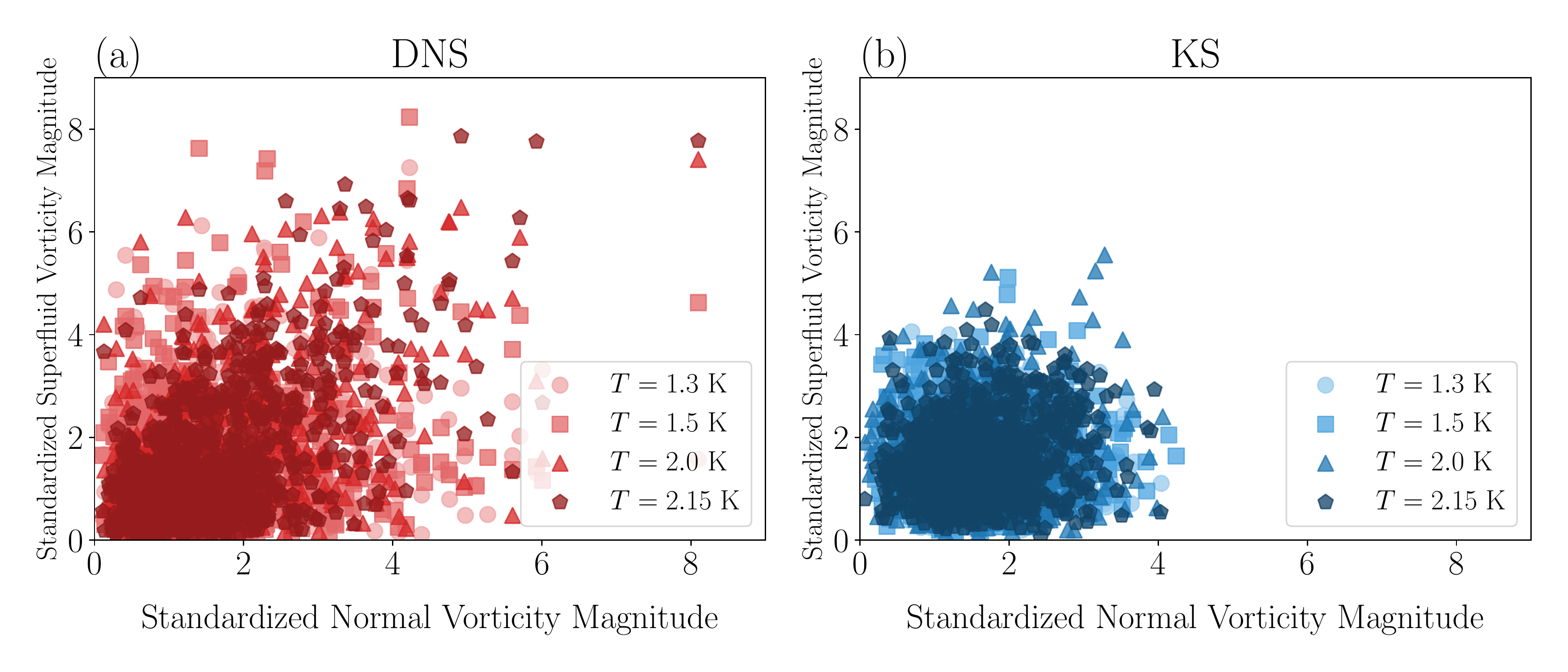}
\caption{Scatter diagrams of the standardized normal vorticity against standardized superfluid vorticity magnitude for both the (a) DNS and (b) KS simulations. Different colors and symbols indicate different temperatures. The data is a representation of a sample of spatial data at a fixed time in statistical steady state conditions. \label{fig:scatter_vort}}
\end{center}
\end{figure*}

\begin{figure}[htp!]
\begin{center}
\includegraphics[width=0.9\columnwidth]{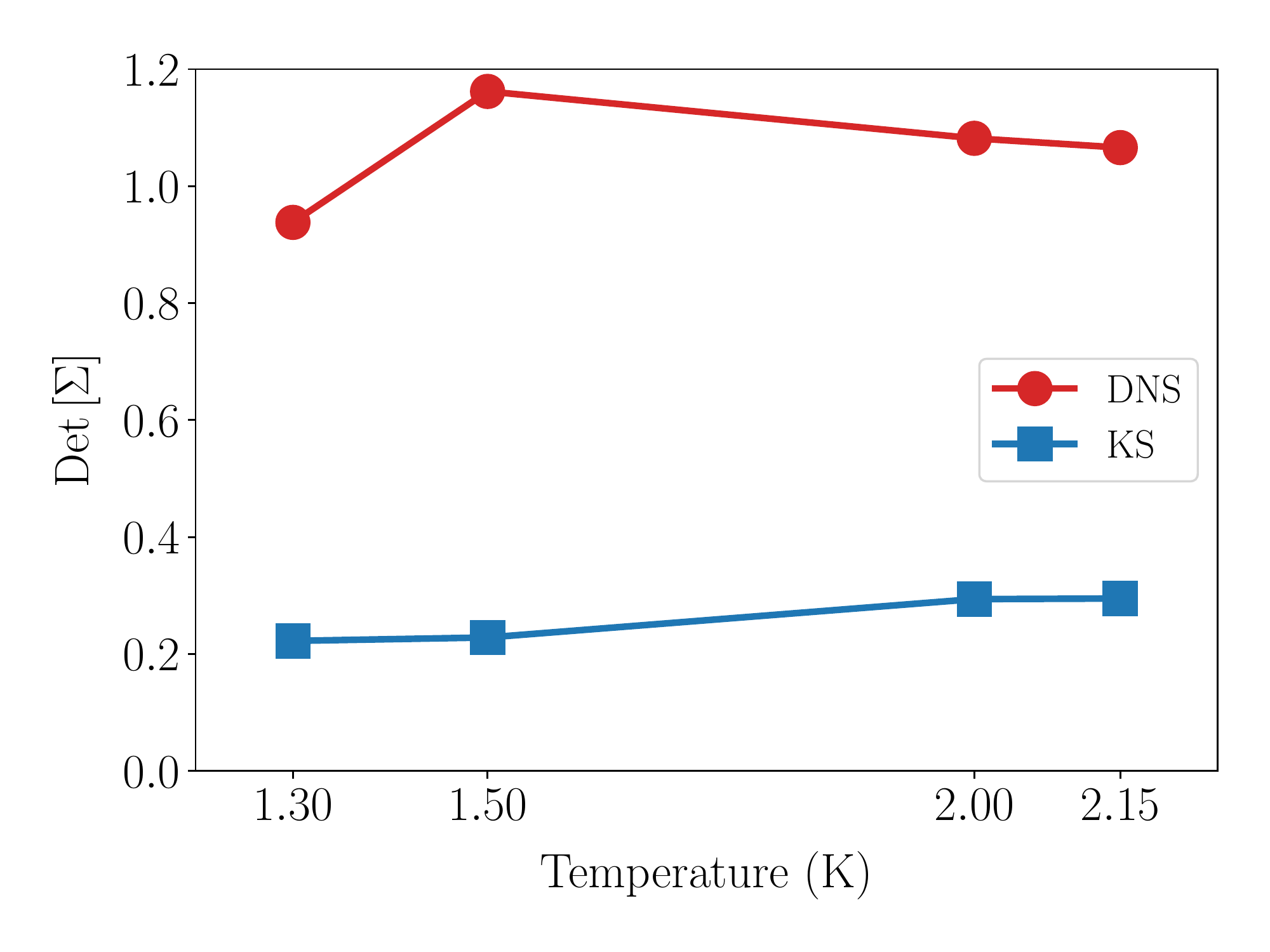}
\caption{The determinant of the covariance matrix $\Sigma$ of the normal and superfluid vorticity data versus temperature. DNS data is plotted as red filled circles, with the KS data as blue filled squares. \label{fig:det}}
\end{center}
\end{figure}

To observe specific patterns in the vorticity correlation we plot scatter diagrams of a sub-sample of the standardized vorticity fields in Fig.~\ref{fig:scatter_vort}. We observe that the DNS data is more susceptible for large vorticity extremes in both components probably related to the fact that the DNS normal fluid flow contains more intense vortical structures when compared to the KS flow. This can be visually ascertained from the enhanced polarization of vortex filaments in Fig.~\ref{fig:tangle}. Interestingly, the extremes in vorticity for both normal and superfluid components do not appear to be particularly correlated between each other with only a slight tendency observed in both the DNS and KS data. For the KS flow, the variability is reduced and more locally centralized around the bulk normalized values. It appears that lower temperatures result in increased variability of the vorticity values. To quantify this spread, in Fig.~\ref{fig:det} we plot the determinant of the $2\times 2$ covariance matrix $\Sigma$. Here, the components of the covariance matrix $\Sigma_{i,j} = {\rm Cov}[X_i,X_j]$ for $i,j=1,2$ and $X_1  = \bar{\omega}_n$ and $X_2 = \bar{\omega}_s$ are the variances and covariance of the normal and superfluid vorticity fields. We observe that computed values of ${\rm Det}\left[\Sigma\right]$ do not appear temperature dependent with the values approximately constant across all temperatures. (Large values of ${\rm Det}\left[\Sigma\right]$ would indicate strong variability of the data.)  However, we do observe increased variability of the vorticity fields normalized by their respective variances. This is likely a consequence of the more significant coherent structures in the DNS flow leading to a larger prevalence of extreme vorticity values in both vorticity fields.

To summarize, we observe clear locking of the velocity and vorticity fields that is enhanced with temperature (via increased mutual friction), that is more prominent in the velocity fields. The correlation occurs at scales larger than the inter-vortex spacing $\ell$, with rapid decoherence of the two-fluid locking at scales below $\ell$. Moreover, we observe evidence that the locking is more significant in the DNS data, which could be the result of the stronger local polarization but could also be related do to the stationary normal fluid state we have used.

Having established the role temperature plays on the correlation of the vorticity fields, we now progress to studying the dynamics of the pressure field. This is an important quantity due to the use of pressure measurements in experiments to infer properties of the underlying velocity fields.

\subsection{Pressure Dynamics in Two Fluid Turbulence}\label{sec:pressure}
\begin{figure*}[t!]
\begin{center}
\includegraphics[width=0.9\textwidth]{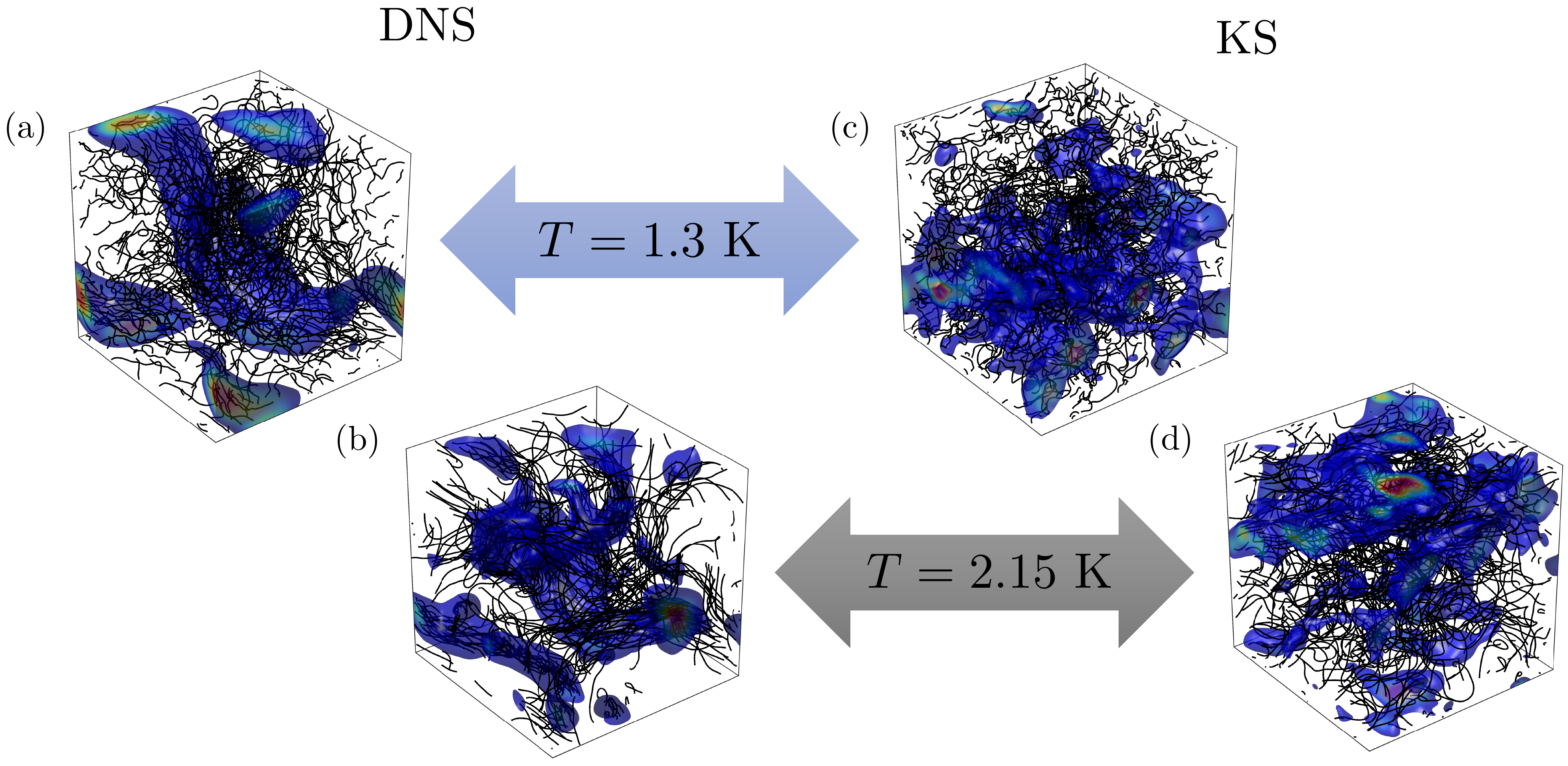}
\caption{Physical space plots of the superfluid vortex filaments of the DNS at (a) $T=1.3~\rm K$, (b) $T=2.15~\rm K$ and KS at (c) $T=1.3~\rm K$, (d) $T=2.15~\rm K$ with overlap of pressure iso-surfaces at the $-1\sigma$-level of the standardized pressure statistics. The negative pressure is situated around large sections of polarization of the superfluid vortex lines.\label{fig:tangle}}
\end{center}
\end{figure*}

In this subsection, we examine the turbulent signatures arising in the resulting two-fluid pressure field. From here on in, our velocity and vorticity fields are filtered with the length scale $l_f=2\ell$ with the pressure computed thereafter. We first examine qualitatively the structure of the vortex tangle and the resulting superfluid pressure fields. Fig.~\ref{fig:tangle} displays snapshots of the vortex tangle for both types of normal fluid drive at the two extreme temperatures of $T=1.3$ and $2.15~{\rm K}$ in statistical steady state conditions. We note that the DNS tangle is more structured, with more defined large-scale superfluid vortical structures (bundles) than in the KS tangle. This is mimicking the structures of the respective normal fluid where phase correlations of modes defining the DNS normal fluid velocity field leads to distinct large-scale coherent structures in the form of vorticity worms highlighted by the pressure iso-surface. In the case of the KS simulations, these phase correlations between Fourier modes are absent and lead to less definitive structures even though the normal fluid energy spectrum remains Kolmogorov $k^{-5/3}$.  We also note that temperature dependence leads to two independent phenomena. First, the mean vortex line curvature is in general larger at lower temperatures due to the presence of Kelvin waves~\cite{svistunov_superfluid_1995}, which become naturally suppressed at higher temperatures by the stronger mutual friction. Second, and more importantly, we observe that in both lower temperature simulations there is a visible increase in the vortex line density away from the highlighted low pressure regions (visualized as pressure iso-surfaces at the $-1\sigma$-level of the standardized pressure statistics). As alluded to earlier, the low pressure regions are associated with the polarized component of the vortex line density, $L_\parallel$, see Eq.~\eqref{eq:Ldecomp}, whose fraction increases with temperature, and hence, we see a picture consistent with the quantitative data presented in Fig.~\ref{fig:polarization} that the level of polarization of the tangle increases with temperature.

\begin{figure*}[htp!]
\begin{center}
\includegraphics[width=\textwidth]{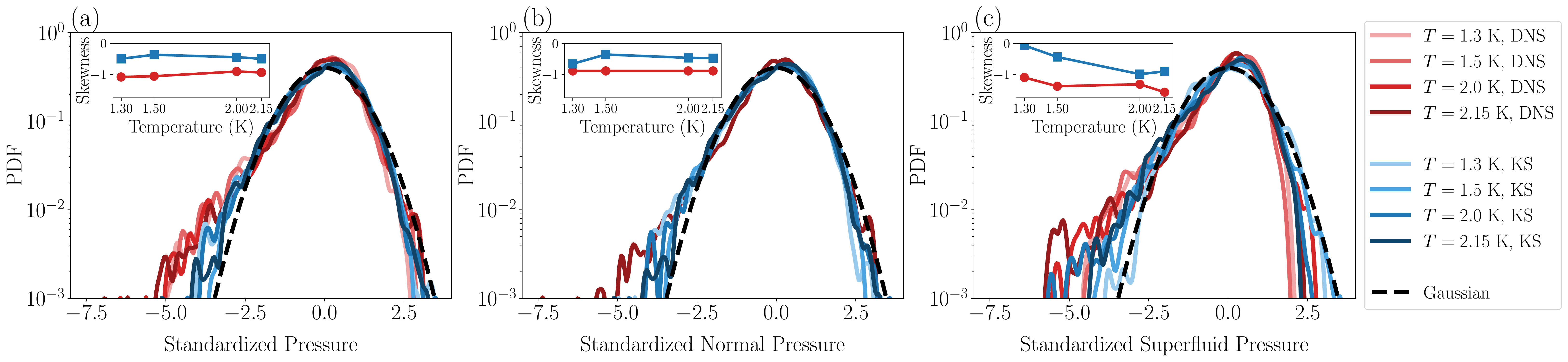}
\caption{Normalized distribution of the fluid pressure $P$ (a) with additional plots of the specific contributions arising from the normal (b) and superfluid (c) fluid components in log-linear coordinates. Colors indicate different temperatures and the DNS and KS simulations respectively. The black dashed curve is of the standard Gaussian distribution. Inserts show the distribution skewness $\left\langle \left( (P-\langle P\rangle)/\sigma_P\right)^3\right\rangle$ versus temperature. \label{fig:pressure_pdf}}
\end{center}
\end{figure*}

The role temperature plays in the pressure dynamics of two-fluid turbulence is most easily understood from studying estimates~\cite{silverman_density_1986} of the Probability Density Functions (PDFs) of the pressure and vorticity fields. Fig.~\ref{fig:pressure_pdf} shows normalized distributions of (a) the total pressure field, alongside the separate contributions arising from (b) the normal and (c) superfluid fluid components as defined in Eq.~\eqref{eq:pressure}. Recall, that the pressure field is dependent on the ratio of fluid densities, hence at low temperatures we expect the pressure field to be dominated by the superfluid contribution and vice-versa at high temperatures.  In both sets of simulations, we observe deviations away from Gaussianity (pure Gaussian behavior is dictated by the black dashed curve), at all temperatures, with a prevalence of introducing more extreme negative pressures. A characterization also observed experimentally in Ref.~\cite{rusaouen_detection_2017}. As we would expect with an imposed normal fluid, the temperature dependence in the pressure PDFs arises due to variation in the density of the superfluid component. This is exasperated at higher temperatures by the structure of the vortex line tangle becoming increasingly locked and more polarized towards the intense vortical regions of the normal fluid flow, more obvious in the DNS data, increasing the non-Gaussian behavior. Consequently, our results suggest that as temperature increases the deviations from Gaussianity become larger, a picture which is not inconsistent with Ref.~\cite{rusaouen_detection_2017}.  The inserts of Fig.~\ref{fig:pressure_pdf} show the computed skewness $\left\langle\left((P-\langle P \rangle)/\sigma_P\right)^3\right\rangle$ of the pressure PDFs. Temperature dependence of the skewness is not expected in the normal pressure contribution due to the one-way coupling of our model, however we do observe consistently more negative values in the DNS simulations and a clear temperature dependence of the superfluid pressure contribution. This is mainly masked in the overall pressure field due to the two-fluid density ratio leading to smaller contributions of the superfluid component at high temperatures. In an attempt to further quantify this process of non-Gaussianity, we compute the Kullback-Leibler (KL) divergence~\cite{mackay_information_2003}: a measure of relative entropy (distance) between two PDFs. The KL divergence between two probability distributions $Q_1$ and $Q_2$ is defined as
\begin{align*}
D_{KL}(Q_1||Q_2) = \int_{-\infty}^\infty Q_1(x)\log\left(\frac{Q_1(x)}{Q_2(x)} \right) \ dx.
\end{align*}

\begin{figure}
  \begin{center}
  \includegraphics[width=0.9\columnwidth]{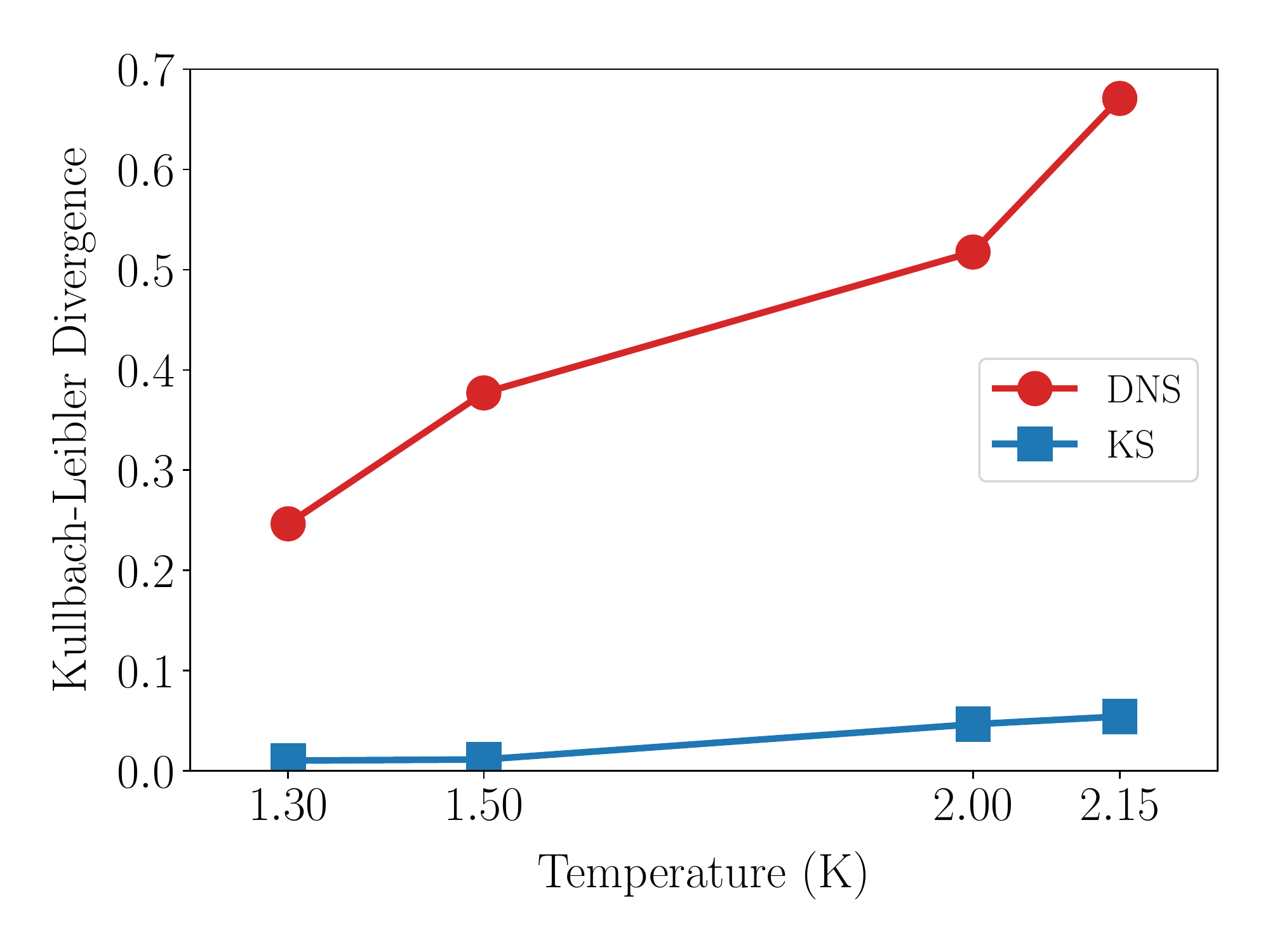}
  \caption{The Kullback-Leibler divergence between the normalized pressure field and the standardized Gaussian distribution.\label{fig:pressure_KLdiv}}
  \end{center}
  \end{figure}
In Fig.~\ref{fig:pressure_KLdiv}, we plot the KL divergence of the measured normalized pressure distribution against the standardized normal distribution. The KL divergence will give zero if the two distributions are identical and provide a positive value that quantifies the difference between the distributions.  We observe an increase of the KL divergence with respect to temperature that is significantly more pronounced in the DNS simulations. Again, this reinforces our hypothesis that the presence of more defined coherent structures is leading to increased non-Gaussianity of the pressure and vorticity fields. This provides additional evidence to the conclusions of Ref.~\cite{rusaouen_detection_2017} where negative pressure extremes where attributed to the presence of coherent structures in the superfluid flow. A natural question arises whether this is connected to the observation of turbulence intermittency in finite temperature superfluids (observed deviations from Kolmogorov scalings in high-order velocity structure functions) that is still in debate as to the extent intermittency occurs~\cite{rusaouen_intermittency_2017}.

\begin{figure*}[htp!]
\begin{center}
\includegraphics[width=0.9\textwidth]{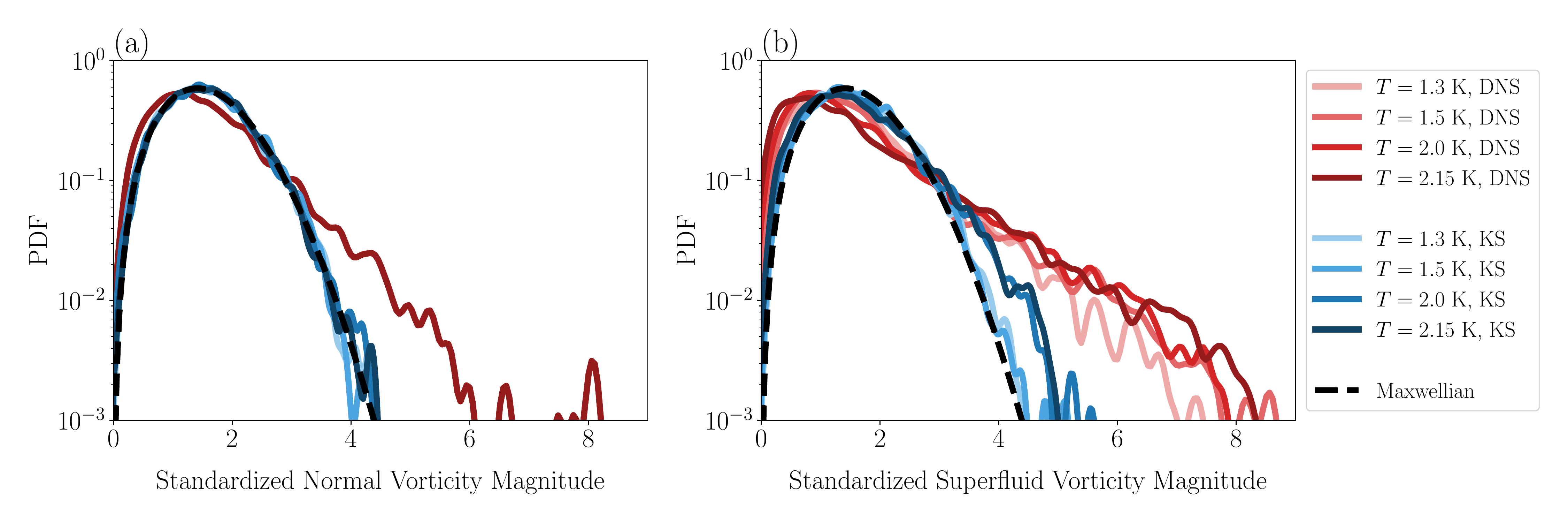}

\caption{Normalized distribution of the normal (a) and superfluid (b) vorticity magnitudes. Colors indicate different temperatures and the DNS and KS simulations respectively.  The black dashed curve is of the standardized Maxwell-Boltzmann distribution that corresponds to Gaussian statistics of the field components.  \label{fig:vorticity_pdf}}
\end{center}
\end{figure*}

As negative pressure extremes are commonly used as a proxy to identify intense superfluid vorticity events, we would expect to observe similar characteristics of the pressure PDFs in distributions of the superfluid vorticity field. These are displayed in Fig.~\ref{fig:vorticity_pdf} and fully explains the temperature dependent dynamics of the pressure field described above. In particular, differences between the pressure fields in the DNS and KS model simulations are due to marked differences in the normal fluid vorticity fields, which the superfluid component locks to. In the case of the static DNS normal fluid data, the normal fluid velocity statistics are identical across all temperatures (as they should due to the static flow only being rescaled by a constant factor), indicated by all PDFs situated on top of one another. This is not the case for the KS model where temporal differences are observed. As very pronounced vortical structure are present in the DNS flow as opposed to the KS flow, these significantly drive the corresponding superfluid flow away from pure Gaussian statistics. For the magnitude of a 3D vector, Gaussian behavior in each component is represented by the Maxwell-Boltzmann distribution, which the unstructured KS superfluid vorticity field more or less follows. Even without these structures in the KS flow, we still observe small but significant deviations from Gaussianity in the superfluid vorticity field (more pronounced at higher temperatures), suggesting that coherent superfluid vortex bundles do not just arise through the presence of coherent structures in the normal fluid, but can arise through the natural nonlinear interactions between the quantized vortices themselves. However, we must conclude that the presence of such structures in the normal fluid significantly enhance this effect, as seen in the DNS simulations. We observe a highly skewed superfluid vorticity distribution with a long temperature dependent tail. All evidence points to the coherent vortical structures driving the presence of extreme vorticity values and the deviation from Gaussian statistics of the field that in turn generates localized regions of low pressure as seen in Figs.~\ref{fig:tangle} and~\ref{fig:pressure_pdf}. 

\begin{figure*}[htp!]
\begin{center}
\includegraphics[width=0.8\textwidth]{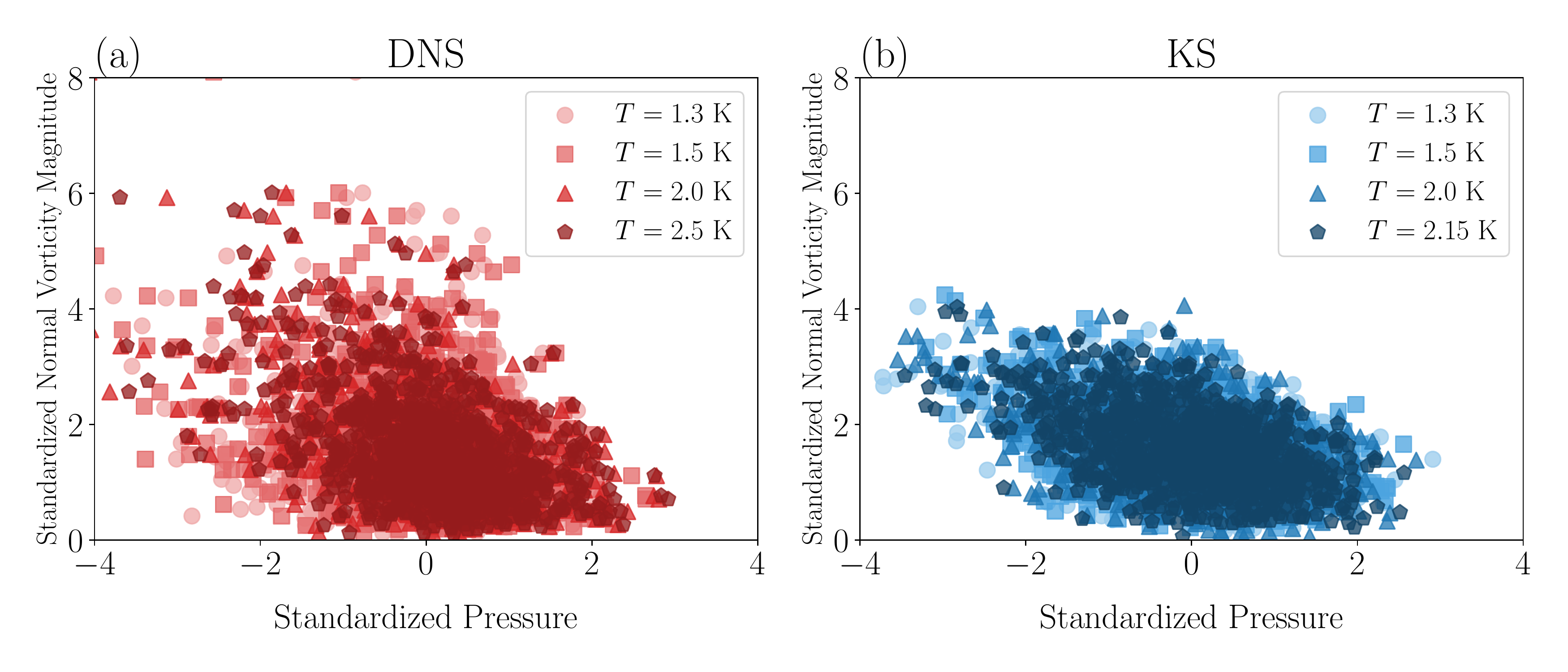}
\caption{Scatter diagrams of the standardized pressure against standardized normal vorticity magnitude for both the (a) DNS and (b) KS simulations. Different colors and symbols indicate different temperatures. The data is a representation of a sample of spatial data at a fixed time in statistical steady state conditions. \label{fig:scatter_press_vort_n}}
\end{center}
\end{figure*}

\begin{figure*}[htp!]
\begin{center}
\includegraphics[width=0.8\textwidth]{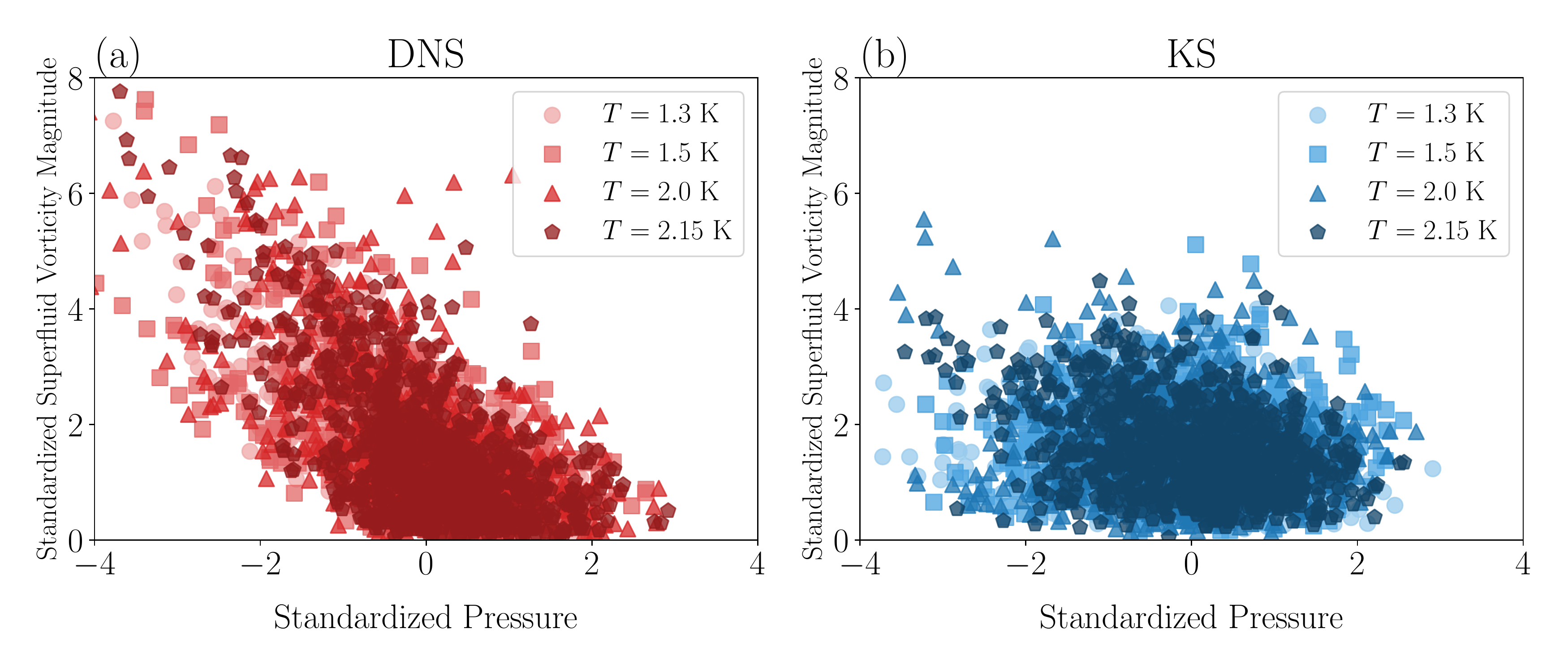}
\caption{Scatter diagrams of the standardized pressure against standardized superfluid vorticity magnitude for both the (a) DNS and (b) KS simulations. Different colors and symbols indicate different temperatures. The data is a representation of a sample of spatial data at a fixed time in statistical steady state conditions. \label{fig:scatter_press_vort_s}}
\end{center}
\end{figure*}

To complete the picture, Figs.~\ref{fig:scatter_press_vort_n} and~\ref{fig:scatter_press_vort_s} present scatter plots of values of the pressure field against the normal and superfluid vorticity fields respectively. We observe a significant negative trend between the superfluid vorticity and pressure particularly for DNS, across all temperatures, which agrees with what was reported in our previous study in Ref.~\cite{laurie_coarse-grained_2020} for zero temperature numerical data. Surprisingly this is less obvious in the case of KS flow, particularly for observing any negative correlation of the pressure with respect to the superfluid vorticity magnitude. We argue that this is likely due to the of lack large coherent structures within the normal fluid KS drive that prevents the formation of strongly localized vortex bundles within the superfluid. This is supported by Fig.~\ref{fig:polarization} where the KS flow appears to generate more random orientated vortex lines that can be visually observed from the vortex line tangle in Fig.~\ref{fig:tangle}.  Moreover, we notice very little dependence in the shape of the scatter-plot data with temperature, even though the vorticity and pressure PDFs (see Figs.~\ref{fig:pressure_pdf} and~\ref{fig:vorticity_pdf}) indicate otherwise. What we can conclude is that the range over which the correlation data is spread may not be temperature dependent, but its internal distribution may be so. This is likely a consequence of, or sensitivity to, the physical structure of the turbulence than the relative ratios between the normal and superfluid components. 

\subsection{The Pressure Spectrum of Superfluid Turbulence}
\noindent Kivotides~\etal~\cite{kivotides_quantum_2001} argued that the 1D pressure spectrum $P_k$ of superfluid turbulence (defined in Eq.~\eqref{eq:pressure}) should scale as $\propto k^{-2}$ due to the localization of superfluid vorticity in quantized vortex filaments, and thus there is a macroscopic quantum signature beyond the classical Kolmogorov theory of turbulence (where the pressure spectrum is $\propto k^{-7/3}$). 
At high temperatures one would expect the pressure field to be dominated by the normal fluid's contribution leading to a $k^{-7/3}$ pressure spectrum signature. However, at low temperatures, where the superfluid density dominates, i.e. below temperatures of $1.5~\rm K$ one may expect to see a transition to the $k^{-2}$ scaling. Indeed, the results of Ref.~\cite{kivotides_quantum_2001} were produced at $T=1.3~{\rm K}$ which indicated such behavior.

While experimentalists are pioneering a new generation of measurement probes which may be able to measure the flow at scales close to the inter-vortex spacing~\cite{salort_micro-cantilever_2011}, scales below this remain inaccessible. Thus, we investigate the 1D pressure spectrum $P_k$ of the unfiltered velocity fields. We define the 1D pressure spectrum $P_k$ as the spectral distribution of the square integrated pressure:
\begin{align}\label{eq:pressure}
\int_{0}^{\infty} P_k \ dk=\frac{1}{{D^3}} \int\left(\frac{P}{\rho}\right)^{2}\ d{\bf x}.
\end{align}

\begin{figure*}[htp!]
\begin{center}
\includegraphics[width=\textwidth]{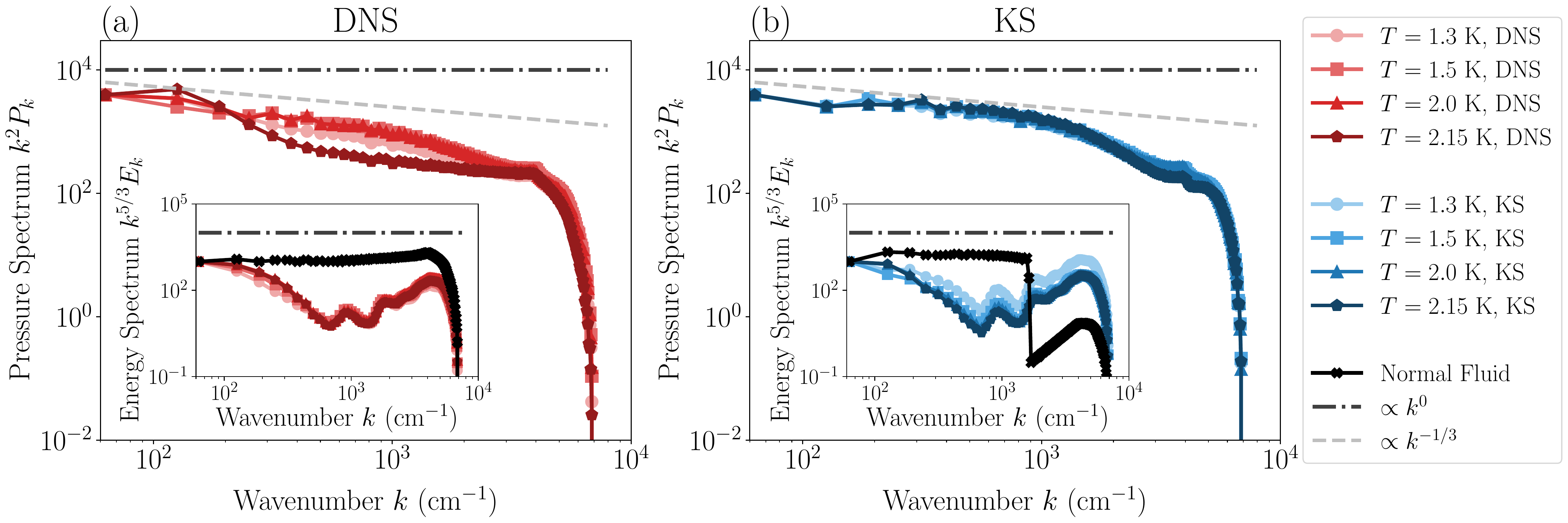}
\caption{(Main) Compensated 1D pressure spectra $k^2P_k$ for the (a) DNS and (b) KS simulations. The pressure spectra are compensated by the Kivotides~\etal quantum pressure spectrum prediction of $P_k \propto k^2$~\cite{kivotides_quantum_2001}. The classical Kolmogorov pressure spectrum of $P_k\propto k^{-7/3}$ is plotted by the light gray dashed line. In all cases, the pressure spectra are computed from the non-coarse-grained velocity fields ${\bf v}_n$ and ${\bf v}_s$.  (Inserts) Compensated 1D energy spectra of the superfluid (colored) and normal (black) velocity fields for each normal flow type respectively. Energy spectra are compensated by the classical Kolmogorov power-law scaling of $E_k\propto k^{-5/3}$. All pressure and energy spectra are additionally shifted vertically to coincide at $k_{\rm min}$.\label{fig:pressure_spectrum}}
\end{center}
\end{figure*}

We present the unfiltered 1D pressure and kinetic energy spectra in Fig.~\ref{fig:pressure_spectrum}. We choose to display the computed spectra from the non-coarse-grained fields due to the argument that the quantum pressure signature is related to the discreteness of the superfluid vorticity field. The main panels display the $k^2$-compensated non-coarse-grained pressure field, with the insets showing the 1D energy spectra $E_k$ (defined through $\int_0^\infty E_k\ dk  = (1/D^3)\int \rho v^2/2 \ d{\bf x}$) computed from the non-coarse-grained velocity fields. Given the very small difference between the two power-law scalings, and the noise present in our numerical data, it is hard to form a definitive opinion. However, we believe there is some evidence that the pressure spectra in the KS simulations is consistent with a $k^{-2}$ scaling (within the range $10^2~{\rm cm}^{-1}<k<10^3 ~{\rm cm}^{-1}$) across all temperatures considered. In contrast, the DNS simulations are more consistent with the K41 scaling, $k^{-7/3}$, again irrespective of temperature.  In performing this analysis, we also compared the pressure spectra produced from the filtered velocity fields (not presented). We were unable to distinguish between the two pressure scalings and saw a lack of power-law behavior due to the effect of small-scale coarse-graining process impacting the inertial range. The energy spectra insets show the clear $k^{-5/3}$-scaling of the normal fluid drive of both the DNS and KS simulations. For the superfluid velocity energy spectra, this is not so clear, but there is evidence of a $k^{-5/3}$-scaling in a small region around an intermediate wavenumber of $10^3~{\rm cm}^{-1}$. We suspect that the lack of a wide agreement is due to the statistical averaging being taken over only one snapshot on a relatively small grid of $128\times128\time 128$ points and the lack of temporal dynamics in the normal fluid drives preventing a true mixing of the superfluid flow. Although not definitively conclusive, it may well be that the pressure spectrum can be used as an indirect measure of the structure of the superfluid vortex tangle. For instance, when the superfluid vortex tangle is relatively unstructured, i.e. when the unpolarized vortex line component dominates $L \approx L_\times$, one may expect a $k^{-2}$ pressure spectrum signature of quantum turbulence to be observed. This could be relevant to superfluid counterflow experiments~\cite{babuin_quantum_2012}, or those where ultra-quantum turbulence is generated through mechanical agitation or pulse charge injection~\cite{walmsley_dynamics_2014}. However, if the polarized component becomes appreciable then one can expect to observe the classical scaling. It would be interesting to revisit these calculations in simulation (or experiments) within the $T=0~{\rm K}$ limit to test if the pressure spectrum can indeed be used as a tool to discriminate between the ultra-quantum and quasi-classical regimes.

As a final comment, the lack of any distinguishable power-law behavior in the pressure spectrum computed using the coarse-grained velocity fields could be problematic for experimentalists as we are frequently using the analogy that the filtering scale $l_f$ is a proxy for the size of an experimental probe. Our results indicate the pressure spectrum of superfluid turbulence may be inaccessible experimentally without probes capable of probing scales much smaller than the inter-vortex spacing.

\section{Discussion}

The purpose of this study was to perform an in depth examination of finite-temperature superfluid turbulence, using the VFM, in regard to the behavior between the externally excited normal fluid and the resulting superfluid flow. We have observed that there is significant increase of vorticity locking at higher temperatures leading to stronger correlations between the two fluid components. This means that any normal fluid characteristics, such as non-Gaussianity of the flow field and the presence of coherent vortical structures are more readily transferred to the superfluid component at higher temperatures. Furthermore, as the KS flow is intrinsically Gaussian, we also show evidence that deviations from Gaussianity can naturally manifest itself directly in the superfluid component which again appears more prominent at higher temperatures. Signatures of which are passed to the pressure field and can thus be detected from analysis of the pressure distribution. However, we are unable to provide conclusive evidence that the measurement of the spatial pressure spectrum can be used to distinguish between classical and quantum turbulent signature as proposed in Ref.~\cite{kivotides_quantum_2001}. Instead, we suggest this spectra may be used to probe the underlying structure of the turbulent tangle.  With that being said, we are only performing simulations that encode a one-way coupling from the normal to superfluid flow, and thus our results, especially at low temperatures, neglect a significant normal fluid feedback from the dominant superfluid flow. Recent advances have been made in this direction~\cite{galantucci_new_2020} with the development of a new high-performance two-way coupled finite-temperature superfluid numerical code utilizing the VFM that in the future may be used to verify our results presented here in more realistic scenarios.


\begin{thebibliography}{0}%
\makeatletter
\providecommand \@ifxundefined [1]{%
 \@ifx{#1\undefined}
}%
\providecommand \@ifnum [1]{%
 \ifnum #1\expandafter \@firstoftwo
 \else \expandafter \@secondoftwo
 \fi
}%
\providecommand \@ifx [1]{%
 \ifx #1\expandafter \@firstoftwo
 \else \expandafter \@secondoftwo
 \fi
}%
\providecommand \natexlab [1]{#1}%
\providecommand \enquote  [1]{``#1''}%
\providecommand \bibnamefont  [1]{#1}%
\providecommand \bibfnamefont [1]{#1}%
\providecommand \citenamefont [1]{#1}%
\providecommand \href@noop [0]{\@secondoftwo}%
\providecommand \href [0]{\begingroup \@sanitize@url \@href}%
\providecommand \@href[1]{\@@startlink{#1}\@@href}%
\providecommand \@@href[1]{\endgroup#1\@@endlink}%
\providecommand \@sanitize@url [0]{\catcode `\\12\catcode `\$12\catcode `\&12\catcode `\#12\catcode `\^12\catcode `\_12\catcode `\%12\relax}%
\providecommand \@@startlink[1]{}%
\providecommand \@@endlink[0]{}%
\providecommand \url  [0]{\begingroup\@sanitize@url \@url }%
\providecommand \@url [1]{\endgroup\@href {#1}{\urlprefix }}%
\providecommand \urlprefix  [0]{URL }%
\providecommand \Eprint [0]{\href }%
\providecommand \doibase [0]{http://dx.doi.org/}%
\providecommand \selectlanguage [0]{\@gobble}%
\providecommand \bibinfo  [0]{\@secondoftwo}%
\providecommand \bibfield  [0]{\@secondoftwo}%
\providecommand \translation [1]{[#1]}%
\providecommand \BibitemOpen [0]{}%
\providecommand \bibitemStop [0]{}%
\providecommand \bibitemNoStop [0]{.\EOS\space}%
\providecommand \EOS [0]{\spacefactor3000\relax}%
\providecommand \BibitemShut  [1]{\csname bibitem#1\endcsname}%
\let\auto@bib@innerbib\@empty
\end{thebibliography}%


\begin{thebibliography}{10}

\bibitem{maurer_local_1998}
J.~Maurer and P.~Tabeling, ``Local investigation of superfluid turbulence,''
  {\em EPL (Europhysics Letters)}, vol.~43, pp.~29--34, July 1998.

\bibitem{walmsley_dynamics_2014}
P.~Walmsley, D.~Zmeev, F.~Pakpour, and A.~Golov, ``Dynamics of quantum
  turbulence of different spectra,'' {\em Proceedings of the National Academy
  of Sciences}, vol.~111, pp.~4691--4698, Mar. 2014.

\bibitem{andersson_superfluid_2007}
N.~Andersson, T.~Sidery, and G.~L. Comer, ``Superfluid neutron star
  turbulence,'' {\em Monthly Notices of the Royal Astronomical Society},
  vol.~381, pp.~747--756, Sept. 2007.

\bibitem{berezhiani_theory_2015}
L.~Berezhiani and J.~Khoury, ``Theory of dark matter superfluidity,'' {\em
  Physical Review D}, vol.~92, p.~103510, Nov. 2015.
\newblock Publisher: American Physical Society.

\bibitem{chesler_holographic_2013}
P.~M. Chesler, H.~Liu, and A.~Adams, ``Holographic {Vortex} {Liquids} and
  {Superfluid} {Turbulence},'' {\em Science}, vol.~341, pp.~368--372, July
  2013.

\bibitem{adams_holographic_2014}
A.~Adams, P.~M. Chesler, and H.~Liu, ``Holographic {Turbulence},'' {\em
  Physical Review Letters}, vol.~112, p.~151602, Apr. 2014.

\bibitem{tisza_theory_1947}
L.~Tisza, ``The {Theory} of {Liquid} {Helium},'' {\em Physical Review},
  vol.~72, pp.~838--854, Nov. 1947.

\bibitem{landau_theory_1941}
L.~Landau, ``Theory of the {Superfluidity} of {Helium} {II},'' {\em Physical
  Review}, vol.~60, pp.~356--358, Aug. 1941.

\bibitem{landau_theory_1949}
L.~Landau, ``On the {Theory} of {Superfluidity},'' {\em Physical Review},
  vol.~75, pp.~884--885, Mar. 1949.

\bibitem{skrbek_four_2000}
L.~Skrbek, J.~J. Niemela, and R.~J. Donnelly, ``Four {Regimes} of {Decaying}
  {Grid} {Turbulence} in a {Finite} {Channel},'' {\em Physical Review Letters},
  vol.~85, pp.~2973--2976, Oct. 2000.

\bibitem{babuin_quantum_2012}
S.~Babuin, M.~Stammeier, E.~Varga, M.~Rotter, and L.~Skrbek, ``Quantum
  turbulence of bellows-driven {4He} superflow: {Steady} state,'' {\em Physical
  Review B}, vol.~86, p.~134515, Oct. 2012.

\bibitem{rousset_superfluid_2014}
B.~Rousset, P.~Bonnay, P.~Diribarne, A.~Girard, J.~M. Poncet, E.~Herbert,
  J.~Salort, C.~Baudet, B.~Castaing, L.~Chevillard, F.~Daviaud, B.~Dubrulle,
  Y.~Gagne, M.~Gibert, B.~H{\'e}bral, T.~Lehner, P.-E. Roche, B.~Saint-Michel,
  and M.~B. Mardion, ``Superfluid high {REynolds} von {K{\'a}rm{\'a}n}
  experiment,'' {\em Review of Scientific Instruments}, vol.~85, p.~103908,
  Oct. 2014.

\bibitem{rusaouen_intermittency_2017}
E.~Rusaouen, B.~Chabaud, J.~Salort, and P.-E. Roche, ``Intermittency of quantum
  turbulence with superfluid fractions from 0\% to 96\%,'' {\em Physics of
  Fluids}, vol.~29, p.~105108, Oct. 2017.

\bibitem{rusaouen_detection_2017}
E.~Rusaouen, B.~Rousset, and P.-E. Roche, ``Detection of vortex coherent
  structures in superfluid turbulence,'' {\em EPL (Europhysics Letters)},
  vol.~118, no.~1, p.~14005, 2017.

\bibitem{paoletti_velocity_2008}
M.~S. Paoletti, M.~E. Fisher, K.~R. Sreenivasan, and D.~P. Lathrop, ``Velocity
  {Statistics} {Distinguish} {Quantum} {Turbulence} from {Classical}
  {Turbulence},'' {\em Physical Review Letters}, vol.~101, p.~154501, Oct.
  2008.

\bibitem{gao_statistical_2017}
J.~Gao, E.~Varga, W.~Guo, and W.~F. Vinen, ``Statistical {Measurement} of
  {Counterflow} {Turbulence} in {Superfluid} {Helium}-4 {Using} {He2}
  {Tracer}-{Line} {Tracking} {Technique},'' {\em Journal of Low Temperature
  Physics}, vol.~187, pp.~490--496, June 2017.

\bibitem{vinen_mutual_1957-2}
W.~F. Vinen, ``Mutual {Friction} in a {Heat} {Current} in {Liquid} {Helium}
  {II}. {I}. {Experiments} on {Steady} {Heat} {Currents},'' {\em Proceedings of
  the Royal Society of London. Series A. Mathematical and Physical Sciences},
  vol.~240, pp.~114--127, Apr. 1957.

\bibitem{vinen_mutual_1957-1}
W.~F. Vinen, ``Mutual {Friction} in a {Heat} {Current} in {Liquid} {Helium}
  {II}. {II}. {Experiments} on {Transient} {Effects},'' {\em Proceedings of the
  Royal Society of London. Series A. Mathematical and Physical Sciences},
  vol.~240, pp.~128--143, Apr. 1957.

\bibitem{vinen_mutual_1957}
W.~F. Vinen, ``Mutual {Friction} in a {Heat} {Current} in {Liquid} {Helium}
  {II}. {III}. {Theory} of the {Mutual} {Friction},'' {\em Proceedings of the
  Royal Society of London. Series A. Mathematical and Physical Sciences},
  vol.~242, pp.~493--515, Nov. 1957.

\bibitem{vinen_mutual_1958}
W.~F. Vinen, ``Mutual {Friction} in a {Heat} {Current} in {Liquid} {Helium}
  {II}. {IV}. {Critical} {Heat} {Currents} in {Wide} {Channels},'' {\em
  Proceedings of the Royal Society of London. Series A. Mathematical and
  Physical Sciences}, vol.~243, pp.~400--413, Jan. 1958.

\bibitem{salort_energy_2012}
J.~Salort, B.~Chabaud, E.~L{\'e}v{\^e}que, and P.-E. Roche, ``Energy cascade
  and the four-fifths law in superfluid turbulence,'' {\em EPL (Europhysics
  Letters)}, vol.~97, p.~34006, Feb. 2012.

\bibitem{morris_vortex_2008}
K.~Morris, J.~Koplik, and D.~W.~I. Rouson, ``Vortex {Locking} in {Direct}
  {Numerical} {Simulations} of {Quantum} {Turbulence},'' {\em Physical Review
  Letters}, vol.~101, p.~015301, July 2008.

\bibitem{laurie_coarse-grained_2020}
J.~Laurie and A.~W. Baggaley, ``Coarse-grained pressure dynamics in superfluid
  turbulence,'' {\em Physical Review Fluids}, vol.~5, p.~014603, Jan. 2020.

\bibitem{hall_rotation_1956}
H.~E. Hall and W.~F. Vinen, ``The {Rotation} of {Liquid} {Helium} {II}. {II}.
  {The} {Theory} of {Mutual} {Friction} in {Uniformly} {Rotating} {Helium}
  {II},'' {\em Proceedings of the Royal Society of London. Series A.
  Mathematical and Physical Sciences}, vol.~238, pp.~215--234, Dec. 1956.

\bibitem{bekarevich_phenomenological_1961}
I.~Bekarevich and I.~Khalatnikov, ``Phenomenological derivation of the
  equations of vortex motion in {He} {II},'' {\em Journal of Experimental and
  Theoretical Physics}, vol.~13, pp.~643--646, 1961.

\bibitem{melotte_transition_1998}
D.~J. Melotte and C.~F. Barenghi, ``Transition to {Normal} {Fluid} {Turbulence}
  in {Helium} {II},'' {\em Physical Review Letters}, vol.~80, pp.~4181--4184,
  May 1998.

\bibitem{barenghi_regimes_2016}
C.~F. Barenghi, Y.~A. Sergeev, and A.~W. Baggaley, ``Regimes of turbulence
  without an energy cascade,'' {\em Scientific Reports}, vol.~6, p.~35701, Oct.
  2016.
\newblock Number: 1 Publisher: Nature Publishing Group.

\bibitem{schwarz_three-dimensional_1985}
K.~W. Schwarz, ``Three-dimensional vortex dynamics in superfluid {He4}:
  {Line}-line and line-boundary interactions,'' {\em Physical Review B},
  vol.~31, pp.~5782--5804, May 1985.

\bibitem{schwarz_three-dimensional_1988}
K.~W. Schwarz, ``Three-dimensional vortex dynamics in superfluid {He4}:
  {Homogeneous} superfluid turbulence,'' {\em Physical Review B}, vol.~38,
  pp.~2398--2417, Aug. 1988.

\bibitem{saffman_vortex_1992}
P.~G. Saffman, {\em Vortex {Dynamics}}.
\newblock Cambridge University Press, 1992.

\bibitem{baggaley_tree_2012}
A.~W. Baggaley and C.~F. Barenghi, ``Tree {Method} for {Quantum} {Vortex}
  {Dynamics},'' {\em Journal of Low Temperature Physics}, vol.~166, pp.~3--20,
  Jan. 2012.

\bibitem{baggaley_sensitivity_2012}
A.~W. Baggaley, ``The {Sensitivity} of the {Vortex} {Filament} {Method} to
  {Different} {Reconnection} {Models},'' {\em Journal of Low Temperature
  Physics}, vol.~168, pp.~18--30, July 2012.

\bibitem{kivotides_spreading_2011}
D.~Kivotides, ``Spreading of superfluid vorticity clouds in normal-fluid
  turbulence,'' {\em Journal of Fluid Mechanics}, vol.~668, pp.~58--75, 2011.

\bibitem{yui_three-dimensional_2018}
S.~Yui, M.~Tsubota, and H.~Kobayashi, ``Three-{Dimensional} {Coupled}
  {Dynamics} of the {Two}-{Fluid} {Model} in {Superfluid}
  \${\textasciicircum}\{4\}{\textbackslash}mathrm\{{He}\}\$: {Deformed}
  {Velocity} {Profile} of {Normal} {Fluid} in {Thermal} {Counterflow},'' {\em
  Physical Review Letters}, vol.~120, p.~155301, Apr. 2018.
\newblock Publisher: American Physical Society.

\bibitem{galantucci_new_2020}
L.~Galantucci, A.~W. Baggaley, C.~F. Barenghi, and G.~Krstulovic, ``A new
  self-consistent approach of quantum turbulence in superfluid helium,'' {\em
  The European Physical Journal Plus}, vol.~135, p.~547, July 2020.

\bibitem{adachi_steady-state_2010}
H.~Adachi, S.~Fujiyama, and M.~Tsubota, ``Steady-state counterflow quantum
  turbulence: {Simulation} of vortex filaments using the full {Biot}-{Savart}
  law,'' {\em Physical Review B}, vol.~81, p.~104511, Mar. 2010.

\bibitem{baggaley_vortex-density_2012}
A.~W. Baggaley, J.~Laurie, and C.~F. Barenghi, ``Vortex-{Density}
  {Fluctuations}, {Energy} {Spectra}, and {Vortical} {Regions} in {Superfluid}
  {Turbulence},'' {\em Physical Review Letters}, vol.~109, p.~205304, Nov.
  2012.

\bibitem{kondaurova_structure_2014}
L.~Kondaurova, V.~L'vov, A.~Pomyalov, and I.~Procaccia, ``Structure of a
  quantum vortex tangle in {4He} counterflow turbulence,'' {\em Physical Review
  B}, vol.~89, p.~014502, Jan. 2014.

\bibitem{smith_decay_1993}
M.~R. Smith, R.~J. Donnelly, N.~Goldenfeld, and W.~F. Vinen, ``Decay of
  vorticity in homogeneous turbulence,'' {\em Physical Review Letters},
  vol.~71, pp.~2583--2586, Oct. 1993.
\newblock Publisher: American Physical Society.

\bibitem{sherwin-robson_local_2015}
L.~K. Sherwin-Robson, C.~F. Barenghi, and A.~W. Baggaley, ``Local and nonlocal
  dynamics in superfluid turbulence,'' {\em Physical Review B}, vol.~91,
  p.~104517, Mar. 2015.

\bibitem{osborne_one-particle_2005}
D.~R. Osborne, J.~C. Vassilicos, and J.~D. Haigh, ``One-particle two-time
  diffusion in three-dimensional homogeneous isotropic turbulence,'' {\em
  Physics of Fluids}, vol.~17, p.~035104, Mar. 2005.
\newblock Publisher: American Institute of Physics.

\bibitem{she_intermittent_1990}
Z.-S. She, E.~Jackson, and S.~A. Orszag, ``Intermittent vortex structures in
  homogeneous isotropic turbulence,'' {\em Nature}, vol.~344, pp.~226--228,
  Mar. 1990.

\bibitem{li_public_2008}
Y.~Li, E.~Perlman, M.~Wan, Y.~Yang, C.~Meneveau, R.~Burns, S.~Chen, A.~Szalay,
  and G.~Eyink, ``A public turbulence database cluster and applications to
  study {Lagrangian} evolution of velocity increments in turbulence,'' {\em
  Journal of Turbulence}, p.~N31, Jan. 2008.

\bibitem{roche_vortex_2008}
P.-E. Roche and C.~F. Barenghi, ``Vortex spectrum in superfluid turbulence:
  {Interpretation} of a recent experiment,'' {\em EPL (Europhysics Letters)},
  vol.~81, p.~36002, Jan. 2008.
\newblock Publisher: IOP Publishing.

\bibitem{krause_mean-field_2016}
F.~Krause and K.-H. R{\"a}dler, {\em Mean-{Field} {Magnetohydrodynamics} and
  {Dynamo} {Theory}}.
\newblock Elsevier, Jan. 2016.

\bibitem{svistunov_superfluid_1995}
B.~V. Svistunov, ``Superfluid turbulence in the low-temperature limit,'' {\em
  Physical Review B}, vol.~52, pp.~3647--3653, Aug. 1995.

\bibitem{silverman_density_1986}
B.~W. Silverman, {\em Density {Estimation} for {Statistics} and {Data}
  {Analysis}}.
\newblock CRC Press, Apr. 1986.

\bibitem{kivotides_quantum_2001}
D.~Kivotides, J.~C. Vassilicos, C.~F. Barenghi, M.~A.~I. Khan, and D.~C.
  Samuels, ``Quantum {Signature} of {Superfluid} {Turbulence},'' {\em Physical
  Review Letters}, vol.~87, p.~275302, Dec. 2001.

\bibitem{salort_micro-cantilever_2011}
J.~Salort, A.~Monfardini, and P.-E. Roche, ``Micro-{Cantilever} {Anemometer}
  for {Cryogenic} {Helium},'' {\em Journal of Physics: Conference Series},
  vol.~318, p.~092030, Dec. 2011.
\newblock Publisher: IOP Publishing.


\bibitem{volovik_classical_2003}
G.E.~Volovik, ``Classical and Quantum Regimes of Superfluid Turbulence,'' {\em Journal of Experimental and Theoretical Physics Letters}, vol.~78, p.~533, Nov. 2003.

\bibitem{monaghan_smoothed_1992} J.~J. Monaghan, ``Smoothed Particle Hydrodynamics,'' {\em  Annual Review  of Astronomy and Astrophysics}, vol.~30, p.~543, Sept. 1992.

\bibitem{mackay_information_2003} D.~J.~C. Mackay, ``Information Theory, Inference, and Learning Algorithms''
\newblock Cambridge University Press, 2003.

\end{thebibliography}
\end{document}